\documentclass[reprint,amsmath,amssymb, superscriptaddress, aps,prb]{revtex4-2}
\usepackage{bbm}
\usepackage{float}
\usepackage{comment}
\usepackage{tabularx}
\usepackage[utf8]{inputenc}
\usepackage[T2A,T1]{fontenc}
\usepackage[english]{babel}
\usepackage{color}
\usepackage{ulem}
\usepackage{xcolor}
\usepackage{hyperref}
\usepackage{soul}
\definecolor{light_green}{HTML}{bdffc4}
\definecolor{light_blue}{HTML}{bdc9ff}
\definecolor{light_orange}{HTML}{ffebbd}
\usepackage{xfrac}
\usepackage{gensymb}
\usepackage{braket}
\usepackage[caption=false]{subfig}
\usepackage{graphicx}
\usepackage{dcolumn}
\usepackage{bm}

\begin{document}
\title{Resonant subspace approximation for photonic crystal slabs}
\author{Nikolay A. Gippius}
\affiliation{Skolkovo Institute of Science and Technology, Bolshoy Boulevard 30, bld. 1, Moscow 121205, Russia}
\author{Ilia M. Fradkin}
\email{I.Fradkin@skoltech.ru}
\affiliation{Skolkovo Institute of Science and Technology, Bolshoy Boulevard 30, bld. 1, Moscow 121205, Russia}
\affiliation{Moscow Institute of Physics and Technology, Institutskiy pereulok 9, Moscow Region 141701, Russia}
\author{Natalia S. Salakhova}
\affiliation{Skolkovo Institute of Science and Technology, Bolshoy Boulevard 30, bld. 1, Moscow 121205, Russia}
\author{Sergey A. Dyakov}
\affiliation{Skolkovo Institute of Science and Technology, Bolshoy Boulevard 30, bld. 1, Moscow 121205, Russia}
\date{\today}
\begin{abstract}
Resonant approximations are indispensable for the interpretation and efficient modeling of photonic crystal slabs, yet most of them describe an eigenmode as a pole in the complex energy plane at a fixed set of the remaining parameters. Tracking such poles and their hybridization across the Brillouin zone or upon variation of structural parameters is labor-intensive and severely limits the use of resonant approximations in band-structure calculations and structural optimization. Here we introduce a resonant subspace approximation that treats the photon energy and all other parameters on an equal footing. Considering a wide class of photonic crystal slabs that can be split into two non-resonant parts, we show that their resonances arise solely from the round-trip propagation of coupled Fourier harmonics between these parts, in direct analogy with Fabry--P\'{e}rot and waveguide modes. We generalize the scalar round-trip phase to round-trip and phase matrices, whose smooth dependence on all parameters allows us to project the problem onto a small resonant subspace defined at a single anchor point and to extrapolate it throughout a local region of parameter space of arbitrary dimensionality. As a result, rigorous computations at only a few points suffice to reconstruct the band structure, modal linewidths, complex hybridization, and optical spectra as functions of energy, wavevector, geometric dimensions, or permittivity within seconds. We demonstrate the accuracy and versatility of the approach on a strong hexagonal silicon grating, resolving intricate mode hybridization, symmetry-protected features, and subtle geometry-controlled effects that are hardly accessible to straightforward computations. The method is fast, accurate, and readily extensible, offering a practical route to the exploration, design, and optimization of resonant photonic crystal slabs.
\end{abstract}
\maketitle
\section{Introduction}

Resonance is a cornerstone phenomenon in modern physics. On the one hand, resonant behavior often serves as a signature of an underlying effect. On the other one, it is routinely exploited to bring out specific, useful physical properties. In nanophotonics, optical resonances appear in spectra of various materials~\cite{sehmi2017optimizing} and structures, enabling their characterization through identification of polaritons~\cite{basov2016polaritons,demenev2008kinetics}, Mie resonances in dielectric nanoparticles~\cite{mie1908beitrage,kruk2017functional}, localized plasmons in plasmonic nanoparticles~\cite{mayer2011localized,willets2007localized}, and many others. At the same time, optical modes are actively used to design structures with desired functionalities. In particular, resonances enhance light–matter interaction, provide feedback in lasers~\cite{chua2011low,baba2015biosensing}, increase the sensitivity of biosensors~\cite{baba2015biosensing,mayer2011localized,willets2007localized} and other optical sensors~\cite{gagliardi2010probing}, sharpen the selectivity of optical filters~\cite{lou2022tunable}, and boost nonlinear optical effects~\cite{soljavcic2004enhancement} such as surface-enhanced Raman scattering~\cite{khannanov2021express} and optical bistability~\cite{cowan2003optical}. Resonant meta-atoms make it possible to realize artificial materials with magnetism~\cite{monticone2014quest, o2002photonic, holloway2003double, wheeler2005three, popa2008compact, ginn2012realizing} and chirality~\cite{wang2016optical, fernandez2019new, ciattoni2015nonlocal, andryieuski2010homogenization, baranov2024effective} far exceeding those found in natural media~\cite{fradkin2026quadrupole,rybin2015phase,zhao2009mie,arbabi2015dielectric, kuznetsov2016optically, staude2017metamaterial, kivshar2018all}. Resonant modes also shape the far-field patterns of antennas~\cite{biagioni2012nanoantennas,giannini2011plasmonic,muhlschlegel2005resonant}. Moreover, all types of waveguide modes—which underpin a significant part of modern optics—can be regarded as resonances themselves~\cite{tikhodeev2002,christ2003waveguide,fradkin2019fourier,fradkin2018fourier}.

The fact that a resonant mode dominates the optical response of a structure in its spectral vicinity greatly simplifies the theoretical description: it allows one to drastically reduce the dimension of the solution space without significant loss of accuracy. This reduction not only speeds up calculations but also provides a clear interpretation of the observed phenomena. Consequently, a broad variety of resonant approximations have been developed, ranging from simple toy models and few-mode optical Hamiltonians to more advanced techniques such as temporal coupled-mode theory~\cite{fan_temporal_2003,suh_temporal_2004}, quasinormal mode expansions~\cite{sauvan2013,bai_efficient_2013,yan_rigorous_2017,gras_quasinormal-mode_2019,ge_quasinormal_2013}, and formulations based on the scattering matrix~\cite{gippius_optical_2004,weiss_analytical_2017,weiss2011strong,gippius2010resonant,weiss_how_2018}. Within these frameworks, considerable effort has been devoted to describing the coupling of resonant modes~\cite{bykov_spatiotemporal_2015,bykov_coupled-mode_2017,bykov_coupled_2023,bykov_obtaining_2024,bykov_coupled-mode_2023,muljarov_rigorous_2025,gippius2010resonant,gromyko2023resonant1,gromyko2023resonant2,weiss2011strong} and their response to perturbations~\cite{muljarov_brillouin-wigner_2010,weiss_dark_2016,yang_simple_2015,yan_shape_2020}, as well as to the proper normalization of the modes and the associated notions of mode volume and scalar product~\cite{kristensen_modes_2013,kristensen_normalization_2015,weiss_analytical_2017,sauvan2013,muljarov2016exact,lalanne_mode_2020,paszkiewicz-idzik_scalar_2024}. The central idea common to most methods is to retain only a few modes that capture the overall behavior with good precision, but in some works, complete bases of resonant modes are even constructed~\cite{muljarov2016resonant,doost2014resonant}, potentially enabling solutions of arbitrary accuracy. A more detailed account can be found in the relevant reviews~\cite{lalanne_light_2017,both_resonant_2021,sauvan2022normalization}.

In the present work we focus on photonic crystal slabs—thin, periodically structured layers that are versatile platforms for manipulating light. Most of these structures are designed to operate in a resonant regime and can be viewed as periodically modulated waveguides, giving rise to a band structure of so-called quasiguided modes. Although such modes share similarities with other resonances, they also possess distinct features. For instance, they are conveniently described within a scattering-matrix formalism~\cite{tikhodeev2002}, they are delocalized and can exhibit infinite quality factors, appearing as dark modes or bound states in the continuum~\cite{hsu2016bound}. Furthermore, their spectral line shapes can become markedly non-Lorentzian near Rayleigh anomalies, where new diffraction channels open~\cite{akimov2011optical,gromyko2022resonant}. Existing approaches can describe quasiguided modes well. However, most of them treat an eigenmode as a pole in the complex energy plane, which is adequate for a fixed set of parameters such as the wavevector $k$, material constants, and geometric dimensions (thickness, meta-atom size, etc.). In simple cases one can track the mode frequencies and field distributions (eigenvectors) as smooth functions of these parameters, though this still requires some effort. Yet, in the majority of photonic crystal slabs, one encounters a much more intricate picture of interacting modes, resulting in complex band structures of hybridized states. Manually sorting and matching modes across different parameter sets is so labor-intensive that it practically cancels the benefits of a resonance-based approximation. Naive automated algorithms, on the other hand, cannot reliably follow the mode evolution and therefore do not eliminate the need for manual verification.
This difficulty severely limits the applicability of resonance approximations in problems that require scanning the entire Brillouin zone or varying structural parameters—for example, calculations of spontaneous emission (Purcell factor)~\cite{dyakov2023purcell}, Casimir forces~\cite{salakhova2026casimir}, near-field energy transfer, or any kind of structural optimization. In this context, there is a strong demand for a universal approach that would naturally account for the hybridization of multiple modes and would treat all parameters, including energy, on an equal footing.

Here, we demonstrate such an approach for a wide class of photonic crystal slabs that can be divided into two non-resonant parts. Resonances of the composite structure arise solely from the interaction between these parts, analogous to Fabry–Pérot or waveguide modes formed between two non-resonant reflectors. To describe them, we introduce a generalized round-trip propagation and phase matrices, each of which governs the eigenmodes formation. Their smooth dependence on the structural and illumination parameters enables us to perform a simple extrapolation from an anchor point in parametric space, thereby accurately capturing a set of prominent modes and their hybridization within a local region of parameter space of any dimensionality. We show that rigorous computations at only three neighboring points in $(\omega-k_\parallel)$ space are sufficient to obtain the effective Hamiltonian in a local environment and to accurately reconstruct the band structure, modes dispersion, linewidths, and even complicated hybridization via "smart" extrapolation within a few seconds. Optical spectra are calculated equally efficiently and might be studied as functions not only of photon energy and $k$-vector, but of any other parameters such as geometric dimensions and the permittivities of the constituent materials. The capability to obtain a quantitative and trustful resonant approximation of the structure as a function of its arbitrary parameters within seconds instead of the typical hours and even days in some cases not only simplifies and greatly accelerates the exploration of photonic crystal slab properties, but also has the potential to dramatically facilitate and may be even change the approach to the design and optimization of resonant structures.

\section{Results}
\subsection{Theoretical approach}
\subsubsection{Fabry--P\'{e}rot and waveguide resonances}

\begin{figure*}
    \centering
    \includegraphics[width=0.85\linewidth]{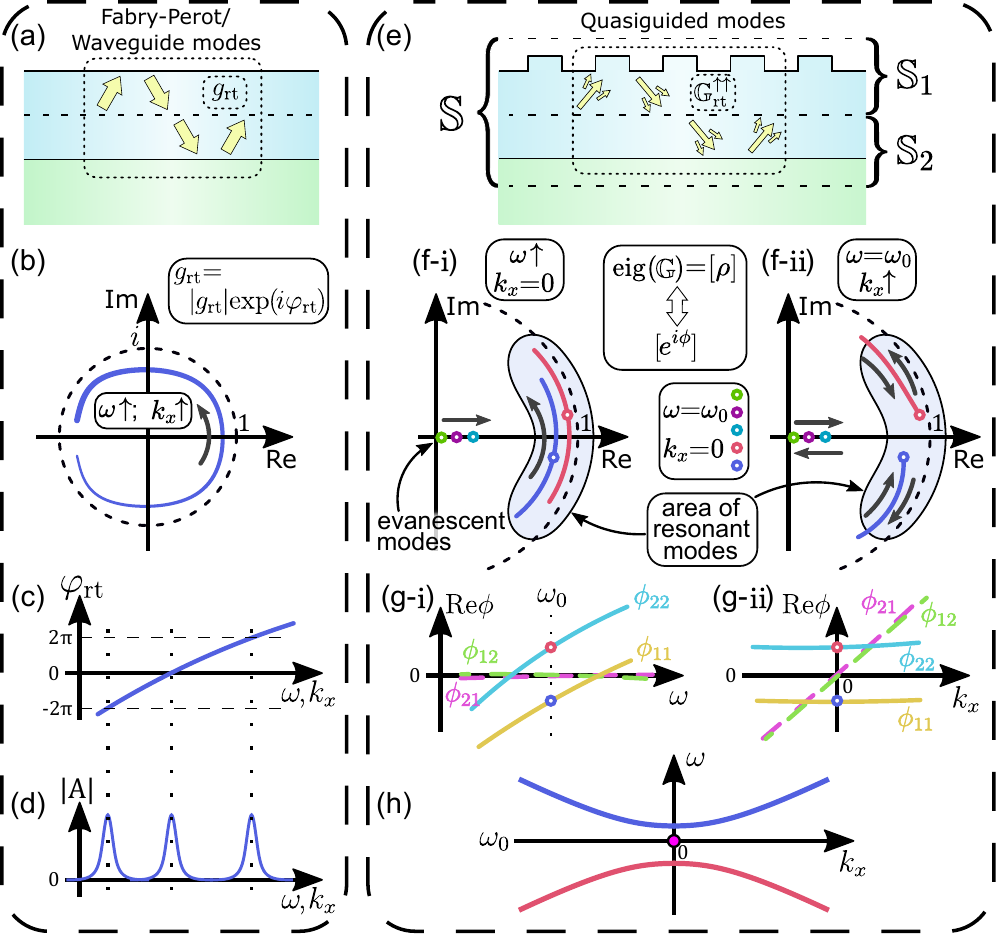}
    \caption{Mechanisms of resonant mode formation in Fabry--P\'{e}rot/slab waveguide structures (a--d) and in photonic crystal slabs (e--h). (a) In a slab waveguide the modes arise from light propagating up and down inside the core. (b) The round-trip amplitude $g_\mathrm{rt}$ generally rotates in the complex plane as a function of frequency or wavevector. (c) The round-trip phase $\varphi_\mathrm{rt}$ is a smooth function of these parameters, and (d) a resonance appears in the spectrum whenever $\varphi_\mathrm{rt}$ is a multiple of $2\pi$. (e) In a photonic crystal slab the modes originate similarly, from the subsequent reflection of light between the upper and lower parts of the core, but light is now a ``bunch'' of coupled Fourier harmonics rather than a plane wave, so the relevant object is the round-trip matrix $\mathbb{G}$. (f-i,f-ii) Its eigenvalues in the complex plane: most stay near zero and correspond to evanescent, non-resonant modes, whereas those approaching unity constitute the resonances (light-blue area). Panels (f-i) and (f-ii) show their evolution with frequency (at $k_x=0$) and with $k_x$ (at $\omega=\omega_0$), respectively; markers indicate the values at the anchor point located at $\Gamma$. (g-i,g-ii) Real part of the corresponding restricted phase matrix, whose smooth parameter dependence (diagonal terms driven mainly by $\omega$, off-diagonal by $k_x$) makes it convenient for extrapolation. (h) The resulting dispersion of the coupled modes, including their hybridization (an anticrossing in this case).}
    \label{fig:fig1}
\end{figure*}

Before turning to periodically modulated photonic crystal slabs, we first discuss the simpler case of a uniform slab waveguide or Fabry--P\'{e}rot resonator formed by several homogeneous layers (see Fig.~\ref{fig:fig1}(a)). The modes of such a structure can have arbitrarily high $Q$-factors and demonstrate different $k$-dependence, but their description is straightforward and well established. The modes originate from light travelling inside the core and undergoing reflections at its boundaries. This process is captured by the complex round-trip propagation coefficient $g_\mathrm{rt}=r_{\uparrow\downarrow}r_{\downarrow\uparrow}e^{2ik_zH}$, where $r_{\uparrow\downarrow}$ and $r_{\downarrow\uparrow}$ are the Fresnel reflection coefficients of the lower and upper interfaces, respectively, $H$ is the thickness of the core, and $k_z$ is the out-of-plane component of the wavevector inside the core. The condition $g_\mathrm{rt}\to 1$ signals the formation of an eigenmode (see Fig.~\ref{fig:fig1}~(b)). It is satisfied when the round-trip phase $\varphi_\mathrm{rt}=\arg g_\mathrm{rt}$ is a multiple of $2\pi$ (see Fig.~\ref{fig:fig1}~(c-d)), which provides a convenient criterion for analysis. Moreover, $g_\mathrm{rt}$ is an explicit and smooth function of all light and structure parameters---frequency, wavevector (see Fig.~\ref{fig:fig1}~(b)), thickness, permittivities, etc.---because neither the reflection coefficients nor the propagation factor are resonant by themselves. Hence a smooth approximation of $g_\mathrm{rt}$ and particularly its phase $\varphi_\mathrm{rt}$ as a function of parameters of our interest can be naturally used to describe the resonant modes and their contribution to various spectra.

\subsubsection{Resonances in photonic crystal slabs}
Compared to a simple slab waveguide, the spectra of photonic crystal slabs are considerably more complex and typically contain at least several narrow, intricately hybridized modes. Nevertheless, in many cases their physical origin remains the same: light reflects from the upper and lower boundaries of a relatively high-index core and acquires the appropriate round-trip phase to form an eigensolution (see Fig.~\ref{fig:fig1}~(e)). At the same time, the in-plane periodical modulation couples different Fourier harmonics, which calls for a somewhat more involved mathematical description, which constitutes the main essence of the study presented below. Notably, although Fig.~\ref{fig:fig1}~(e) depicts a weakly modulated grating for illustrative simplicity, this coupling need not be weak: our approach does not treat the modulation as a perturbation and applies equally to strong gratings---precisely the case considered in the validation below. But before going to development of the numerical approach, let us briefly announce what we expect to obtain by developing the resonant approximation based on the same principle for the photonic crystal slab. Now, we should face a set of eigenvalues of some round trip operator, some of which correspond to the hybrid, resonant modes (see Fig.~\ref{fig:fig1}~(f-i,ii)). They travel in the complex plane differently with a change of different parameters of the problem, such as frequency (panel (f-i)) or $k$-vector (panel (f-ii)).
Notably, hybridized modes generally acquire different quality factors, so that one eigenvalue approaches unity more closely (the narrow, higher-$Q$ mode) while the other stays slightly farther away (the wider, lower-$Q$ mode). Their evolution with $k_x$ is non-monotonic, the eigenvalues first approaching unity and then receding along the same path.
The convenient way to track the parameter-dependence of the eigenvalues is to connect the round-trip matrix with some phase matrix of the modes, whose dependence on the typical parameters is expected to be rather smooth for simple and accurate extrapolation - panels (g-i) for typical $\omega$-dependence and panel (g-ii) for $k_x$-one. Finally, all together, it should allow us to reconstruct dispersion and linewidth of the modes as a function of the considered parameters panel (h) as well as the contribution of the modes to all sort of the optical response and spectra of the structure.

\begin{figure*}
    \centering
    \includegraphics[width=0.8\linewidth]{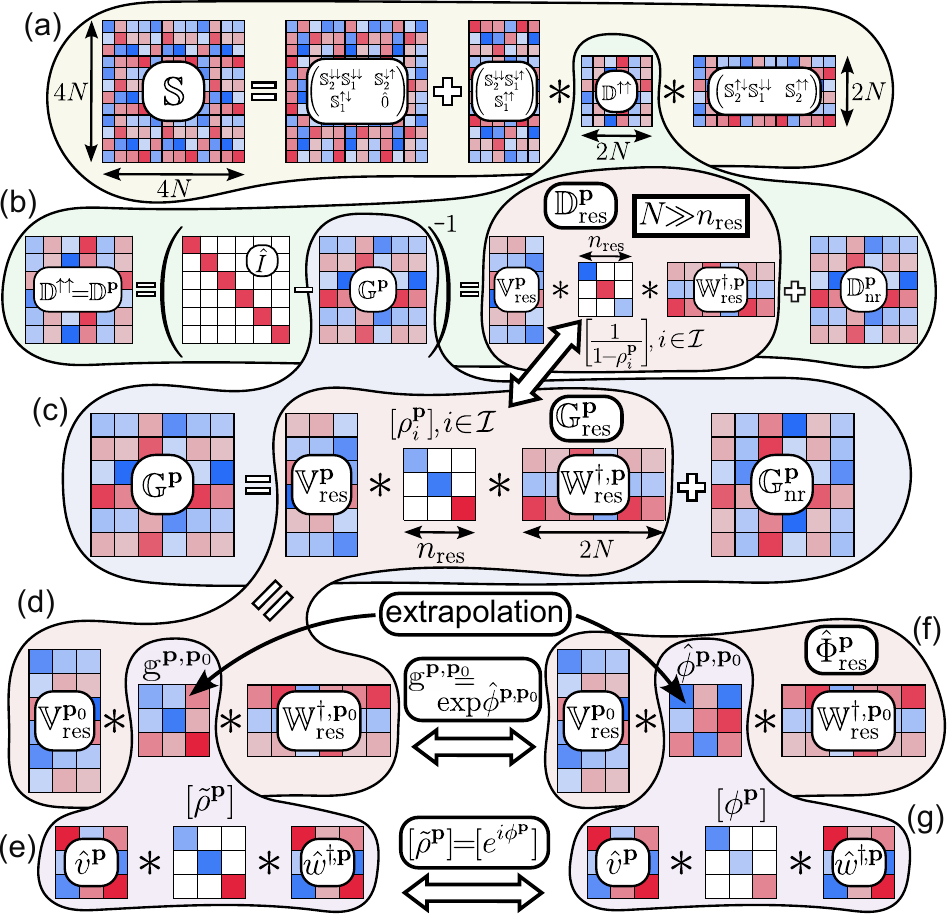}
    \caption{Scheme of the connections between the matrices, illustrating the resonant subspace approach. (a) The resonances of the scattering matrix $\mathbb{S}$ originate from the denominator-like matrix $\mathbb{D}$ (b), which is in turn expressed through the round-trip matrix $\mathbb{G}$ (c). Both $\mathbb{D}$ and $\mathbb{G}$ can be split into resonant and non-resonant parts: knowledge of the resonant eigenvalues $[\rho^\mathbf{p}]_\mathrm{res}$ and eigenvectors $\mathbb{V}^\mathbf{p}_\mathrm{res}$, $\mathbb{W}^\mathbf{p}_\mathrm{res}$ is sufficient to reconstruct $\mathbb{D}_\mathrm{res}$ (b,c). To evaluate $\mathbb{G}_\mathrm{res}^\mathbf{p}$ efficiently at an arbitrary point $\mathbf{p}$, we apply the resonant subspace approximation and express the matrix in the fixed basis of the anchor point $\mathbf{p}_0$. (d) This reduces the large matrix $\mathbb{G}_\mathrm{res}^\mathbf{p}$ to the small restricted round-trip matrix $\mathbbm{g}^{\mathbf{p},\mathbf{p}_0}$, which is easily extrapolated in parameter space and diagonalized for each value of $\mathbf{p}$ (e). (f,g) The same procedure can equivalently be carried out for the restricted phase matrix $\hat{\phi}^{\mathbf{p},\mathbf{p}_0}$, which is uniquely related to $\mathbbm{g}^{\mathbf{p},\mathbf{p}_0}$ but is often better suited to a linear-like extrapolation.}
    \label{fig:fig2}
\end{figure*}

We employ a convenient scattering matrix formalism, which relates the amplitudes of incoming and outgoing from the described layer waves. All scattering matrices in this work are computed with an in-house implementation of the Fourier modal method (FMM)~\cite{tikhodeev2002}, also known as rigorous coupled-wave analysis (RCWA)~\cite{moharam1995}.
A key advantage of scattering matrices is that they allow one to derive the optical properties of a composite structure directly from those of its constituents. In particular, the whole photonic crystal slab of our interest is described by the matrix $\mathbb{S}$, but it can be naturally subdivided into an upper ($\mathbb{S}_1$) and a lower ($\mathbb{S}_2$) part  (see Fig.~\ref{fig:fig1} (e)), which are combined according to the rule $\mathbb{S} = \mathbb{S}_1\otimes\mathbb{S}_2$ (see Appendix~\ref{sec:app_dd} for an equivalent formulation) as follows (see Fig.~\ref{fig:fig2}~(a) for visualization):
\begin{multline}
\mathbb{S}=\begin{pmatrix}
    \mathbb{S}^{\downarrow\downarrow} & \mathbb{S}^{\downarrow\uparrow}\\
    \mathbb{S}^{\uparrow\downarrow} & \mathbb{S}^{\uparrow\uparrow}
\end{pmatrix}
=\\
\begin{pmatrix}
  \mathbb{S}_2^{\downarrow\downarrow}  \mathbb{S}_1^{\downarrow\downarrow}                             &    \mathbb{S}_2^{\downarrow\uparrow}   \\
  \mathbb{S}_1^{\uparrow\downarrow}    &    \hat{0}
\end{pmatrix}
+
\begin{pmatrix}
    \mathbb{S}_2^{\downarrow\downarrow}  \mathbb{S}_1^{\downarrow\uparrow} \\
      \mathbb{S}_1^{\uparrow\uparrow}
\end{pmatrix}
  \mathbb{D}^{\uparrow\uparrow}
\begin{pmatrix}
\mathbb{S}_2^{\uparrow\downarrow} \mathbb{S}_1^{\downarrow\downarrow} &     \mathbb{S}_2^{\uparrow\uparrow}
\end{pmatrix}
,
\label{Eq:Ss_DAA}
\end{multline}
where
\begin{equation}
    \mathbb{D}^{\uparrow\uparrow} = \left(\hat{I}-\mathbb{S}_2^{\uparrow\downarrow}\mathbb{S}_1^{\downarrow\uparrow}\right)^{-1}.
    \label{eq:DAA}
\end{equation}
The arrow superscripts on the sub-matrices $\mathbb{S}^{\downarrow\downarrow}$, $\mathbb{S}^{\downarrow\uparrow}$, $\mathbb{S}^{\uparrow\downarrow}$, $\mathbb{S}^{\uparrow\uparrow}$ indicate the propagation directions of the incoming and outgoing waves associated with the respective blocks. Importantly, each of these blocks, as well as $\mathbb{D}^{\uparrow\uparrow}$, is rather large, of $2N\times2N $ size, where $N$ is the number of Fourier harmonics and a factor of 2 corresponds to two possible polarizations. The required number of Fourier harmonics might strongly depend on the particular structure and materials, but typically it is large enough ($N\gg1$), which is one of the main sources of computational complexity that the developed approach aims to resolve.

We consider a wide class of the photonic crystal slabs for which the isolated upper and lower parts do not possess any intrinsic resonant optical response. Hence $\mathbb{S}_1$ and $\mathbb{S}_2$ are non-resonant and depend smoothly on the problem parameters. The full structure, however, generally supports a variety of quasi-guided modes. Equation~\eqref{Eq:Ss_DAA} reveals that in such case resonances can only arise from the $\mathbb{D}^{\uparrow\uparrow}$ term (see Fig.~\ref{fig:fig2}~(b)). This term embodies a Fabry--P\'{e}rot-like process in which light bounces back and forth between the upper and lower layers. The matrices $\mathbb{S}_1^{\downarrow\uparrow}$ and $\mathbb{S}_2^{\uparrow\downarrow}$ generalize the scalar reflection coefficients $r^{\downarrow\uparrow}$ and $r^{\uparrow\downarrow}$ introduced earlier. Qualitatively, $\mathbb{D}^{\uparrow\uparrow}$ can be interpreted as a geometric series summing all multiple reflections:
\begin{equation}
    \mathbb{D}^{\uparrow\uparrow} = \left(\hat{I}-\mathbb{S}_2^{\uparrow\downarrow}\mathbb{S}_1^{\downarrow\uparrow}\right)^{-1} = \sum_{n=0}^\infty \left(\mathbb{S}_2^{\uparrow\downarrow}\mathbb{S}_1^{\downarrow\uparrow}\right)^n .
\end{equation}

For brevity, we introduce the round-trip propagation matrix and denote it by $\mathbb{G}$, the letter being visually reminiscent of the circular arrow symbol $\reflectbox{\rotatebox{90}{$\circlearrowright$}}$ that illustrates the round-trip propagation:
\begin{equation}
    \mathbb{G}\stackrel{\mathrm{def}}{=}\mathbb{G}_\mathrm{rt}^{\uparrow\uparrow} = \mathbb{S}_2^{\uparrow\downarrow}\mathbb{S}_1^{\downarrow\uparrow}.
\end{equation}
In what follows we drop the superscript ``$\uparrow\uparrow$'' (and define $\mathbb{D}\equiv\mathbb{D}^{\uparrow\uparrow}$) as well as the subscript ``rt'' for the brevity, since derivations for the opposite propagation direction ($\downarrow\downarrow$) are completely analogous.
The round-trip matrix is the sole quantity that governs the resonances of the composite structure and is therefore of central interest. Because it is expressed solely through the smooth, non-resonant matrices of the upper and lower sublayers, it depends smoothly on all parameters as well.
To analyze the resonance structure, we proceed with the spectral decomposition of $\mathbb{G}$, which is well-defined away from exceptional points, where $\mathbb{G}$ remains diagonalizable.
\begin{equation}
    \mathbb{G}\mathbb{V} = \mathbb{V}[ \rho ],
\end{equation}
where $[\rho]=\operatorname{diag}(\rho_1,\rho_2,\dots)$ is the diagonal matrix of eigenvalues and $\mathbb{V}=(\ket{\mathbf{V}_1},\ket{\mathbf{V}_2},\dots)$ is the matrix of associated right eigenvectors. Because $\mathbb{G}$ is non-Hermitian, $\mathbb{V}$ is generally non-unitary ($\mathbb{V}^{-1}\neq\mathbb{V}^\dagger$). It is then convenient to introduce the matrix of left eigenvectors $\mathbb{W}=(\ket{\mathbf{W}_1},\ket{\mathbf{W}_2},\dots)$ such that
\begin{equation}
    \mathbb{V}^{-1} = \mathbb{W}^\dagger = \begin{pmatrix}
        \bra{\mathbf{W}_1} \\ \bra{\mathbf{W}_2} \\ \vdots
    \end{pmatrix}.
\end{equation}
By construction this gives the biorthonormality relation $\braket{\mathbf{W}_i|\mathbf{V}_j} = \delta_{ij}$, and we can write the round-trip matrix as
\begin{equation}
    \mathbb{G} = \mathbb{V} [\rho] \mathbb{W}^\dagger = \sum_i \ket{\mathbf{V}_i} \rho_i \bra{\mathbf{W}_i}.
\end{equation}
From this spectral representation the denominator-like matrix $\mathbb{D}$ follows directly:
\begin{multline}
    \mathbb{D} = (\hat{I}-\mathbb{G})^{-1} = (\hat{I}-\mathbb{V}[\rho]\mathbb{W}^\dagger)^{-1}
    =\\ \mathbb{V}\left[ \frac{1}{1-\rho} \right]\mathbb{W}^\dagger
    = \sum_i \ket{\mathbf{V}_i} \frac{1}{1-\rho_i} \bra{\mathbf{W}_i}.
    \label{Eq:D_sum}
\end{multline}
Equation~\eqref{Eq:D_sum} makes it explicit that a resonance occurs exactly when one of the eigenvalues of the round-trip matrix approaches unity, $\rho_i \approx 1$, and it is namely the desired matrix characterizing eigenmodes of the structure, whose eigenvalues are graphically illustrated in Fig.~\ref{fig:fig1}~(f-i,ii). Hence, analyzing the smooth operator $\mathbb{G}$ paves the way for an accurate description of the resonant matrix $\mathbb{D}$ and, consequently, of the whole scattering matrix $\mathbb{S}$ of interest.

\subsubsection{Resonant subspace}
To simplify the calculations, it is more effective not merely to work with large operators in their eigenbasis, but instead to explicitly separate a relatively small resonant subspace of interest from the non-resonant background, which can be smoothly approximated. Indeed, most terms in Eq.~\ref{Eq:D_sum} correspond to eigenvalues far from the resonance condition and represent the non-resonant background. In particular, some eigenvalues might correspond to truly resonant modes that are nevertheless very far from resonance itself ($\rho$ far from 1, but approaching it for some other values of parameters). Moreover, if the number of Fourier harmonics $N$ is rather large, most of the modes are evanescent, since they correspond to the round-trip decay of the field whose energy lies below the light cone of the core of the waveguide for a given $k_\parallel$ value. Evanescent modes might become resonant with increase of frequency, but until their eigenvalues are close to zero value ($\rho\approx0$), we are not interested in them (see Fig.~\ref{fig:fig1}~(f-i,ii)). We therefore introduce the set of eigenstates whose eigenvalues are close to unity and constitute the resonances within or close to the range of interest:
\begin{equation}
    \mathcal{I} = \{ i \in \{1,…,2N\}\mid \rho_i \text{ is near } 1\}.
\end{equation}

Importantly, the number of elements in this set, which is the number of resonant modes, is typically much smaller than the total number of Fourier harmonics, $|\mathcal{I}|=n_\mathrm{res}\ll N$.
The precise criterion of proximity to unity can be chosen in different ways. One possible way to define the area of interest is graphically depicted by light-blue areas in Fig.~\ref{fig:fig1}~(f-i,ii). Another simple, but yet practical choice is \(|\rho_i-1|<\delta\), where \(\delta\) is a threshold typically taken between 0.1 and 0.9. This allows us to decompose the denominator matrix \(\mathbb{D}\) into a resonant and a non-resonant part (see Fig.~\ref{fig:fig2}~(b)):
\begin{multline}
    \mathbb{D} = \mathbb{D}_\mathrm{res} + \mathbb{D}_\mathrm{nr} \\
    = \sum_{i\in\mathcal{I}} \ket{\mathbf{V}_i}\frac{1}{1-\rho_i}\bra{\mathbf{W}_i}
    + \sum_{i\notin\mathcal{I}} \ket{\mathbf{V}_i}\frac{1}{1-\rho_i}\bra{\mathbf{W}_i}.
\end{multline}

We are primarily interested in the low-rank resonant part  \(\mathbb{D}_\mathrm{res}\) that carries the most valuable information about the modes. The high-rank non-resonant matrix \(\mathbb{D}_\mathrm{nr}\) represents a smooth “background” and does not need to be treated separately. In the end, we shall directly approximate the background of the full scattering matrix \(\mathbb{S}\) without passing through intermediate quantities.
The above definition of \(\mathbb{D}_\mathrm{res}\) based on the eigen-decomposition is illustrative, but it becomes impractical when one wishes to follow the parametric \(\mathbf{p}\)-dependence of the resonances. By \(\mathbf{p}\) we imply the set of parameters such as light energy, wavevector and thickness, permittivities and others. Even a weak coupling between modes causes hybridization, which makes the eigenvectors \(\ket{\mathbf{V}_i^{\mathbf{p}}}\) and \(\bra{\mathbf{W}_i^{\mathbf{p}}}\) rotate rapidly in a narrow avoided-crossing region of parameter space.
Fortunately, \(\mathbb{D}_\mathrm{res}^{\mathbf{p}} \overset{\text{def}}{=} \mathbb{D}_\mathrm{res}(\mathbf{p})\) can be determined at any parametric point \(\mathbf{p}\) in a more convenient way. Consider the associated low-rank resonant part of the round-trip matrix (see Fig.~\ref{fig:fig2}~(c)):
\begin{equation}
    \mathbb{G}_\mathrm{res} = \sum_{i\in\mathcal{I}} \ket{\mathbf{V}_i}\rho_i \bra{\mathbf{W}_i}.
    \label{eq:G_res_spec}
\end{equation}
Given \(\mathbb{G}_\mathrm{res}\), the matrix \(\mathbb{D}_\mathrm{res}\) can be uniquely recovered (and vice versa, see Fig.~\ref{fig:fig2}~(b-c)). Crucially, \(\mathbb{G}_\mathrm{res}\) depends smoothly on the parameters (will be shown below) in practice, unlike \(\mathbb{D}_\mathrm{res}\).
Indeed, \(\mathbb{G}_\mathrm{res}\) can be obtained not only from the spectral representation \eqref{eq:G_res_spec}, but also as the restriction of the full round-trip matrix \(\mathbb{G}\) to the resonant subspace:
\begin{equation}
    \mathbb{G}_\mathrm{res}^\mathbf{p} = \mathbb{P}_\mathrm{res}^\mathbf{p} \mathbb{G}^\mathbf{p} \mathbb{P}_\mathrm{res}^\mathbf{p},
\end{equation}
where \(\mathbb{P}_\mathrm{res}^\mathbf{p}\) is the projector onto the resonant subspace at point \(\mathbf{p}\). The crucial observation is that, although the individual eigenvectors \(\ket{\mathbf{V}_i^\mathbf{p}}\) and \(\ket{\mathbf{W}_i^\mathbf{p}}\) of the resonant modes may vary strongly when modes hybridize, the resonant subspace as a whole — and therefore its projector — remains nearly constant in a local region of parameter space:
\begin{equation}
    \mathbb{P}_\mathrm{res}^\mathbf{p}
    = \sum_{i\in\mathcal{I}} \ket{\mathbf{V}_i^\mathbf{p}}\bra{\mathbf{W}_i^\mathbf{p}}
    = \mathbb{V}_\mathrm{res}^\mathbf{p} \mathbb{W}_\mathrm{res}^{\dagger,\mathbf{p}}
    \approx \mathrm{const},
\end{equation}
where \(\mathbb{V}_\mathrm{res}^\mathbf{p}\) and \(\mathbb{W}_\mathrm{res}^\mathbf{p}\) are the matrices composed of the resonant right and left eigenvectors, respectively.

This assertion is our first approximation; it is not proven but holds in practice because the coupling between the resonant modes and the non-resonant background is weak and depends only weakly on the parameters. Consequently, \(\mathbb{G}_\mathrm{res}^\mathbf{p}\), being the restriction of a smooth matrix to an essentially fixed subspace, is a smooth function of the parameters as well.
The approximation of a nearly constant projector allows us to avoid tracking its evolution. Instead, we evaluate the projector at a chosen anchor point \(\mathbf{p}_0\) and then express the round-trip matrix in that fixed basis (see Fig.~\ref{fig:fig2}~(d)):
\begin{multline}
    \mathbb{G}_\mathrm{res}^\mathbf{p}
    = \mathbb{P}_\mathrm{res}^\mathbf{p} \mathbb{G}^\mathbf{p} \mathbb{P}_\mathrm{res}^\mathbf{p}
    \approx \mathbb{P}_\mathrm{res}^{\mathbf{p}_0} \mathbb{G}^\mathbf{p} \mathbb{P}_\mathrm{res}^{\mathbf{p}_0} \\
    = \mathbb{V}_\mathrm{res}^{\mathbf{p}_0}
      \bigl( \mathbb{W}_\mathrm{res}^{\mathbf{p}_0,\dagger} \mathbb{G}^\mathbf{p} \mathbb{V}_\mathrm{res}^{\mathbf{p}_0} \bigr)
      \mathbb{W}_\mathrm{res}^{\mathbf{p}_0,\dagger} \\
    = \mathbb{V}_\mathrm{res}^{\mathbf{p}_0} \,
      \mathbbm{g}^{\mathbf{p},\mathbf{p}_0} \,
      \mathbb{W}_\mathrm{res}^{\dagger,\mathbf{p}_0}
    = \sum_{i,j\in\mathcal{I}} \ket{\mathbf{V}_i^{\mathbf{p}_0}} \,
      \mathbbm{g}_{ij}^{\mathbf{p},\mathbf{p}_0} \,
      \bra{\mathbf{W}_j^{\mathbf{p}_0}},
\end{multline}
where
\begin{equation}
    \mathbbm{g}_{ij}^{\mathbf{p},\mathbf{p}_0}
    = \braket{\mathbf{W}_i^{\mathbf{p}_0} | \mathbb{G}^\mathbf{p} | \mathbf{V}_j^{\mathbf{p}_0}}
\end{equation}
is the round-trip matrix projected onto the resonant subspace and expressed in the fixed basis of the anchor point \(\mathbf{p}_0\). We refer to \(\mathbbm{g}^{\mathbf{p},\mathbf{p}_0}\) as the \textit{restricted round-trip matrix} (in the \(\mathbf{p}_0\)-basis). It contains the full information about the resonances and their parameter dependence, and has the same rank as the original \(\mathbb{G}_\mathrm{res}\), but in contrast to large $2N\times 2N$-sized matrix \(\mathbb{G}_\mathrm{res}\), its size is only \(n_\mathrm{res}\times n_\mathrm{res}\), which makes it convenient and efficient to handle. Moreover, this formulation of a projection on a fixed basis automatically circumvents the necessity of consistently matching the basis states when different parameter points are treated separately.

The restricted round-trip matrix can always be computed directly as the projection of \(\mathbb{G}^\mathbf{p}\). However, our goal is the reverse: we aim to estimate the small matrix \(\mathbbm{g}^{\mathbf{p},\mathbf{p}_0}\) at any point \(\mathbf{p}\) by a computationally cheap extrapolation (or interpolation), and then reconstruct the full resonant matrix \(\mathbb{G}_\mathrm{res}^\mathbf{p}\) from it (see Fig.~\ref{fig:fig2}~(c-d)).
For example, if we are interested in the frequency dependence, a linear approximation can be obtained from only one additional point \(\mathbbm{g}^{\omega_0+\delta\omega,\omega_0}\) besides the anchor point \(\mathbbm{g}^{\omega_0,\omega_0}\):
\begin{multline}
    \mathbbm{g}^{\omega,\omega_0}
    \approx \mathbbm{g}^{\omega_0,\omega_0}
      + \frac{\mathbbm{g}^{\omega_0+\delta\omega,\omega_0} - \mathbbm{g}^{\omega_0,\omega_0}}{\delta\omega}
        (\omega-\omega_0),
\end{multline}
where the zeroth-order term is, by definition, the diagonal matrix
\(\mathbbm{g}^{\omega_0,\omega_0} = \mathbbm{g}^{\mathbf{p}_0,\mathbf{p}_0} = \operatorname{diag}(\rho_i)\big|_{i\in\mathcal{I}}\). This scheme naturally extends to an arbitrary number of parameters: one simply estimates the first derivatives at the anchor point along each direction by evaluating the matrix at \(N_p\) additional points (the number of parameters of interest) besides the anchor one. If needed, extrapolation can be non-linear or based on any other suitably smooth functions.

Once \(\mathbbm{g}^{\mathbf{p},\mathbf{p}_0}\) has been estimated at a desired point \(\mathbf{p}\), its spectral decomposition is numerically straightforward because the matrix is small (see Fig.~\ref{fig:fig2}~(e)):
\begin{equation}
    \mathbbm{g}^{\mathbf{p},\mathbf{p}_0}
    = \hat{v}^\mathbf{p} \, [\tilde{\rho}^\mathbf{p}] \, \hat{w}^{\mathbf{p},\dagger}
    = \sum_{m} v_{im}^\mathbf{p} \, \tilde{\rho}_m^\mathbf{p} \, w_{jm}^{\mathbf{p},*},
\end{equation}
where \(\hat{v}^\mathbf{p}\) and \(\hat{w}^\mathbf{p}\) are the matrices of right and left eigenvectors of \(\mathbbm{g}^{\mathbf{p},\mathbf{p}_0}\), and \(\tilde{\rho}_m^\mathbf{p}\) are the corresponding eigenvalues. By the $\sim$ sign we emphasize that the corresponding eigenvalue is an approximation, not the true value. From this representation we can also reconstruct the resonant part of the round-trip matrix:
\begin{multline}
    \mathbb{G}_\mathrm{res}^\mathbf{p}
    = \mathbb{V}_\mathrm{res}^{\mathbf{p}_0} \,
      \hat{v}^\mathbf{p} \, [\tilde{\rho}^\mathbf{p}] \,
      \hat{w}^{\dagger,\mathbf{p}} \,
      \mathbb{W}_\mathrm{res}^{\dagger,\mathbf{p}_0} \\
    = \tilde{\mathbb{V}}_\mathrm{res}^{\mathbf{p}} \,
      [\tilde{\rho}^\mathbf{p}] \,
      \tilde{\mathbb{W}}_\mathrm{res}^{\dagger,\mathbf{p}}
    = \sum_{m} \ket{\tilde{V}_m^{\mathbf{p}}} \,
      \tilde{\rho}_m^\mathbf{p} \,
      \bra{\tilde{W}_m^{\mathbf{p}}},
    \label{eq:G_res_final}
\end{multline}
where
\[
    \ket{\tilde{V}_m^{\mathbf{p}}} = \sum_{i\in\mathcal{I}} \ket{\mathbf{V}_i^{\mathbf{p}_0}} \, v_{im}^\mathbf{p},
    \qquad
    \ket{\tilde{W}_m^{\mathbf{p}}} = \sum_{i\in\mathcal{I}} \ket{\mathbf{W}_i^{\mathbf{p}_0}} \, w_{im}^\mathbf{p},
\]
and \(\tilde{\rho}_m^\mathbf{p}\) are the estimated eigenvalues, while \(\ket{\tilde{V}_m^{\mathbf{p}}}\) and \(\ket{\tilde{W}_m^{\mathbf{p}}}\) are the estimated eigenvectors of the resonant states at point \(\mathbf{p}\). In this picture, the new eigenvectors are expressed as superpositions of the eigenvectors at the anchor point \(\mathbf{p}_0\).

\subsubsection{Phase matrix}
The algorithm described so far is already operational, but does not yet account for the fact that the eigenvalues of the round-trip matrix often exhibit an exponential dependence on parameters such as frequency, thickness, etc., causing them to rotate rapidly in the complex plane. In such a situation, for the efficient extrapolation of the restricted round-trip matrix as well as for the convenient introduction of the effective Hamiltonian, it is convenient to introduce the phase matrix of the resonant subspace,
\begin{equation}
    \hat{\Phi}^\mathbf{p}_\mathrm{res} = -i \sum_{i\in\mathcal{I}} \ket{\mathbf{V}^\mathbf{p}_i}
        \ln \rho^\mathbf{p}_i \bra{\mathbf{W}^\mathbf{p}_i},
\end{equation}
where the branch of the logarithm is chosen to ensure continuity when moving to neighboring points.
The matrix \(\hat{\Phi}_\mathrm{res}\) generalizes the scalar round-trip phase \(\varphi_\mathrm{rt}\) to the case of several coupled resonant modes. Note that, unlike \(\varphi_\mathrm{rt}\), the matrix \(\hat{\Phi}_\mathrm{res}\) encodes not only the phase but also the amplitude of the round-trip eigenvalues. This choice keeps the formalism compact by avoiding separate amplitude and phase quantities, at the cost of making the phase matrix complex-valued.
Because \(\hat{\Phi}_\mathrm{res}\) shares the same eigenbasis as \(\mathbb{G}_\mathrm{res}\), the two matrices can be calculated one from the other. In practice it is advantageous to work with the \textit{restricted phase matrix} expressed in the basis of the anchor point \(\mathbf{p}_0\) (see Fig.~\ref{fig:fig2}~(f)):
\begin{equation}
    \hat{\phi}^{\mathbf{p},\mathbf{p}_0}_{ij} = \bra{\mathbf{W}_i^{\mathbf{p}_0}} \hat{\Phi}^\mathbf{p}_\mathrm{res} \ket{\mathbf{V}_j^{\mathbf{p}_0}}.
\end{equation}
The restricted round-trip matrix is then simply the matrix exponential (see Appendix~\ref{sec:app_phase} for derivation) of the restricted phase matrix (see Fig.~\ref{fig:fig2}~(d,f)),
\begin{equation}
    \mathbbm{g}^{\mathbf{p},\mathbf{p}_0} = \exp\!\bigl(i\hat{\phi}^{\mathbf{p},\mathbf{p}_0}\bigr).
\end{equation}
The matrix exponential for a small matrix might be calculated through the spectral decomposition of the phase matrix:
\begin{equation}
    \mathbbm{g}^{\mathbf{p},\mathbf{p}_0} = \hat{v}^\mathbf{p} \bigl[e^{i\phi^\mathbf{p}}\bigr] \hat{w}^{\dagger,\mathbf{p}}
    = \sum_{m} v_{im}^\mathbf{p}\, e^{i\phi_m^\mathbf{p}}\, w_{jm}^{\mathbf{p},*},
\end{equation}
where \(\hat{v}^\mathbf{p}\) and \(\hat{w}^\mathbf{p}\) are the right and left eigenvectors of \(\hat{\phi}^{\mathbf{p},\mathbf{p}_0}\), and \(\phi_m^\mathbf{p}\) are its eigenvalues. The eigenvectors of the round-trip matrix are the same as those of the phase matrix, and the eigenvalues are obtained as (see Fig.~\ref{fig:fig2}~(e,g))
\begin{equation}
    \tilde{\rho}_m^\mathbf{p} = e^{i\phi_m^\mathbf{p}}.
\end{equation}
Importantly, $\hat{\phi}$ is namely the matrix that we have desired to obtain from the very beginning (see Fig.~\ref{fig:fig1}~(g-i,ii)), and now we have derived the connection of this matrix with the eigenvalues of the round-trip matrix. All further derivations remain identical regardless of whether one extrapolates the round-trip matrix directly or determines it through the phase matrix.
The phase-matrix approach described here is merely one convenient way to perform an efficient extrapolation - a linear fit of the phase matrix directly yields an exponential behaviour for the round-trip matrix. Other effective fitting strategies certainly exist and deserve future exploration.
\subsubsection{Resonant expansion for the scattering matrix}
\begin{figure*}
    \centering
    \includegraphics[width=1\linewidth]{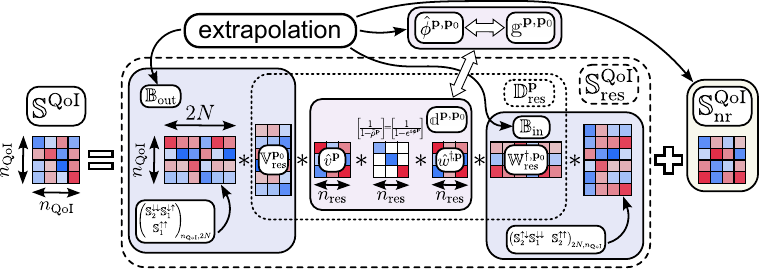}
    \caption{Scheme illustrating the calculation of the reduced scattering matrix $\mathbb{S}^\mathrm{QoI}$, which comprises only a relatively small number of quantities of interest (QoI). To compute it at an arbitrary point in parameter space, one needs a smooth extrapolation of just four matrices: the non-resonant part $\mathbb{S}_\mathrm{nr}^\mathrm{QoI}$, the input and output coupling matrices $\mathbb{B}_\mathrm{in}$ and $\mathbb{B}_\mathrm{out}$, which describe the coupling of the resonances to the far field, and, most importantly, the restricted round-trip matrix $\mathbbm{g}^{\mathbf{p},\mathbf{p}_0}$ (or, equivalently, the phase matrix $\hat{\phi}^{\mathbf{p},\mathbf{p}_0}$) that describes the resonances themselves. The first three are used directly, whereas the last one yields the restricted ``denominator'' matrix $\mathbbm{d}^{\mathbf{p},\mathbf{p}_0}$ at any point in parameter space. The scheme illustrates the connections between these quantities and the ones introduced earlier.}
    \label{fig:fig3}
\end{figure*}

Having obtained the resonant part of the round-trip matrix (Eq.~\ref{eq:G_res_final}), we now derive the corresponding expansion for the full scattering matrix.
As discussed above, once the spectral representation of \(\mathbb{G}_\mathrm{res}\) is known, \(\mathbb{D}_\mathrm{res}\) follows immediately (see Fig.~\ref{fig:fig2}~(b-c)):
\begin{equation}
    \mathbb{D}_\mathrm{res} \approx \tilde{\mathbb{V}}_\mathrm{res}^\mathbf{p}
        \left[\frac{1}{1-\tilde{\rho}}\right]
        \tilde{\mathbb{W}}_\mathrm{res}^{\mathbf{p},\dagger}
    = \sum_i \ket{\tilde{V}_i^\mathbf{p}} \frac{1}{1-\tilde{\rho}_i^\mathbf{p}} \bra{\tilde{W}_i^\mathbf{p}}.
\end{equation}
Substituting this into Eq.~\ref{Eq:Ss_DAA} gives the resonant part of the scattering matrix,
\begin{equation}
    \mathbb{S}_\mathrm{res} = \begin{pmatrix}
            \mathbb{S}_2^{\downarrow\downarrow} \mathbb{S}_1^{\downarrow\uparrow} \\[2pt]
            \mathbb{S}_1^{\uparrow\uparrow}
        \end{pmatrix}\mathbb{V}_\mathrm{res}^{\mathbf{p}_0}
        \hat{v}^\mathbf{p} \left[\frac{1}{1-\tilde{\rho}^\mathbf{p}}\right] \hat{w}^{\dagger,\mathbf{p}}\mathbb{W}_\mathrm{res}^{\dagger,\mathbf{p}_0}
        \begin{pmatrix}
            \mathbb{S}_2^{\uparrow\downarrow} \mathbb{S}_1^{\downarrow\downarrow} & \mathbb{S}_2^{\uparrow\uparrow}
        \end{pmatrix}.
\end{equation}
This matrix has a large $4N\times4N$ size. Nevertheless, we have needed the large number of harmonics only to obtain the valid optical properties of the sublayers and potentially their interaction. At the final stage, we are typically interested only in several quantities of interest ($\mathrm{QoI}$). These QoI are the main channels in most cases, but might actually be any of them. The main idea is that their number is much smaller than the number of Fourier harmonics, $n_\mathrm{QoI}\ll N$ and scattering matrix, $\mathbb{S}^\mathrm{QoI}$, connecting them is also correspondingly small ($n_\mathrm{QoI}\times n_\mathrm{QoI}$). Concurrently, the number of QoI, $n_\mathrm{QoI}$, and the number of resonant modes, $n_\mathrm{res}$, might relate arbitrarily to each other. In this scope, we obtain the following connection of the quantities of interest (see Fig.~\ref{fig:fig3}):
\begin{equation}
    \mathbb{S}^\mathrm{QoI}=\mathbb{S}_\mathrm{res}^\mathrm{QoI}+\mathbb{S}_\mathrm{nr}^\mathrm{QoI} =\mathbb{B}_\mathrm{out}\,
        \mathbbm{d}^{\mathbf{p},\mathbf{p}_0}\mathbb{B}_\mathrm{in}+\mathbb{S}_\mathrm{nr}^\mathrm{QoI},
\end{equation}
where the input and output coupling matrices are
\begin{align}
    \mathbb{B}_\mathrm{out} &=
        \begin{pmatrix}
            \mathbb{S}_2^{\downarrow\downarrow} \mathbb{S}_1^{\downarrow\uparrow} \\[2pt]
            \mathbb{S}_1^{\uparrow\uparrow}
        \end{pmatrix}_{n_\mathrm{QoI},2N}
        \mathbb{V}_\mathrm{res}^{\mathbf{p}_0}, \\
        \mathbbm{d}^{\mathbf{p},\mathbf{p}_0}&= \hat{v}^\mathbf{p} \left[\frac{1}{1-\tilde{\rho}^\mathbf{p}}\right] \hat{w}^{\dagger,\mathbf{p}},\\
    \mathbb{B}_\mathrm{in}  &=
        \mathbb{W}_\mathrm{res}^{\mathbf{p}_0,\dagger}
        \begin{pmatrix}
            \mathbb{S}_2^{\uparrow\downarrow} \mathbb{S}_1^{\downarrow\downarrow} & \mathbb{S}_2^{\uparrow\uparrow}
        \end{pmatrix}_{2N,n_\mathrm{QoI}}.
\end{align}
The matrices \(\mathbb{B}_\mathrm{out}\) and \(\mathbb{B}_\mathrm{in}\) describe the coupling of the resonant part to the external environment. Importantly, these matrices are rather small, of $n_\mathrm{QoI}\times n_\mathrm{res}$ and $n_\mathrm{res}\times n_\mathrm{QoI}$ size, respectively.
The \(\mathbb{B}\) matrices could in principle be recomputed at each point \(\mathbf{p}\). However, even with a limited number of QoI this would still involve the large matrices \(\mathbb{S}_{1,2}\) and \(\mathbb{V}_\mathrm{res}\), and would therefore compromise the computational speedup. To avoid this, we calculate them only at a few points in the immediate vicinity of the anchor point \(\mathbf{p}_0\) and then extrapolate them smoothly, exactly as was done for the restricted round-trip matrix \(\mathbbm{g}^{\mathbf{p},\mathbf{p}_0}\).

The same strategy is applied to the non-resonant part of the scattering matrix. We compute \(\mathbb{S}_\mathrm{nr} = \mathbb{S} - \mathbb{S}_\mathrm{res}\) at the same extrapolation points and then interpolate or extrapolate it as a smooth function of the parameters.
The core of the expression is the restricted resonant part of the denominator matrix $\mathbbm{d}^{\mathbf{p},\mathbf{p}_0}$. As a resonant quantity, it obviously cannot be approximated by any smooth function. However, it is easily obtained from the spectral decomposition of the restricted round-trip matrix $\mathbbm{g}^{\mathbf{p},\mathbf{p}_0}$, which is in turn successfully extrapolated as discussed above.

\subsubsection{From the phase matrix to an effective Hamiltonian}
The procedure described above yields the scattering matrix in a resonance-revealing form. Often, however, one is interested directly in the eigenmodes, their dispersion, and the parameter dependence of their field profiles, rather than in their contribution to some particular optical responses such as transmission/reflection spectra and so on. In such a case it is convenient to develop something like a Hamiltonian operator, whose spectrum would provide us with both eigenvalues and eigenvectors. The most demanded, although not the only, case is the one in which we would like to track the frequency $\omega_\mathrm{res}$ of the modes as a function of all other parameters $\mathbf{p}$. As we will see, the restricted phase matrix \(\hat{\phi}^{(\omega,\mathbf{p}),(\omega_0,\mathbf{p}_0)}\) is an extremely convenient quantity for this purpose, since its frequency dependence is well approximated by linear extrapolation. The resonance condition reads as follows
\begin{equation}
    \hat{\phi}^{\mathbf{p},\mathbf{p}_0}(\omega_\mathrm{res},\omega_0) \Psi_\mathrm{res} = 0,
\end{equation}
where $\Psi_\mathrm{res}$ is the resonant vector of the mode expressed in the basis of the resonant eigenvectors $\mathbb{V}^{(\omega_0,\mathbf{p}_0)}_\mathrm{res}$ at $(\omega_0,\mathbf{p}_0)$ anchor point.
Assuming a linear expansion in frequency around a reference frequency \(\omega_0\), we obtain
\begin{equation}
    \hat{\phi}^{\mathbf{p},\mathbf{p}_0}(\omega_0,\omega_0)\Psi_\mathbf{res}
    + \left.\frac{\partial\hat{\phi}^{\mathbf{p},\mathbf{p}_0}(\omega,\omega_0)}{\partial\omega}\right|_{\omega=\omega_0}(\omega_\mathrm{res}-\omega_0)\Psi_\mathbf{res} = 0,
\end{equation}
which is easily rearranged into the classical eigenvalue problem:
\begin{equation}
    \Bigl[ \omega_0 + \Bigl(\left.\frac{\partial\hat{\phi}^{\mathbf{p},\mathbf{p}_0}(\omega,\omega_0)}{\partial\omega}\right|_{\omega=\omega_0}\Bigr)^{-1}
        \hat{\phi}^{\mathbf{p},\mathbf{p}_0}(\omega_0,\omega_0) \Bigr] \Psi_\mathrm{res} = \omega_\mathrm{res} \Psi_\mathrm{res}.
\end{equation}
Thus we are led to an effective Hamiltonian obtained in a local environment of anchor point $(\omega_0,\mathbf{p}_0)$:
\begin{equation}
    \hat{H}^{(\omega_0,\mathbf{p}_0)}(\mathbf{p}) =     \Bigl[ \omega_0 + \Bigl(\left.\frac{\partial\hat{\phi}^{\mathbf{p},\mathbf{p}_0}(\omega,\omega_0)}{\partial\omega}\right|_{\omega=\omega_0}\Bigr)^{-1}
        \hat{\phi}^{\mathbf{p},\mathbf{p}_0}(\omega_0,\omega_0) \Bigr].
\end{equation}
The dependence on the remaining parameters \(\mathbf{p}\) can be approximated by a simple (e.g., linear) extrapolation as well. Because \(\hat{H}\) is a relatively small \(n_\mathrm{res}\times n_\mathrm{res}\) matrix, its eigenvalue problem can be solved at a very large number of points in a reasonable timescale, enabling easy reconstruction of the full dispersion and hybridization landscape. Most importantly, we are freed from the necessity to compute the coupling of a large number of high-$k$ harmonics that define the accurate optical properties of the structure for each separate point. All the required information was obtained from the characterization of the anchor point and is stored with the phase matrix and its derivative in parametric space. The same strategy might be easily applied not only to the resonant frequency but to any other resonant parameter, such as the slab thickness or the permittivity of a material. Also, if we would like to obtain a more precise result, we can consider a non-linear dependence of the parameter of interest and apply the existing approaches for non-linear Hamiltonians.
The eigenvector \(\Psi_\mathrm{res}\) corresponds to the up-going part of the resonant field at the interface between the upper and lower sublayers.

Finally, we note that a similar Hamiltonian can be formally written in terms of the restricted round-trip matrix \(\mathbbm{g}\). Although a linear fit of \(\mathbbm{g}\) with respect to frequency is typically much less accurate than that of the phase matrix, it might be beneficial for the resonant expansions with respect to some other parameters.

\subsection{Numerical demonstration}
\begin{figure}
    \centering
    \includegraphics[width=\linewidth]{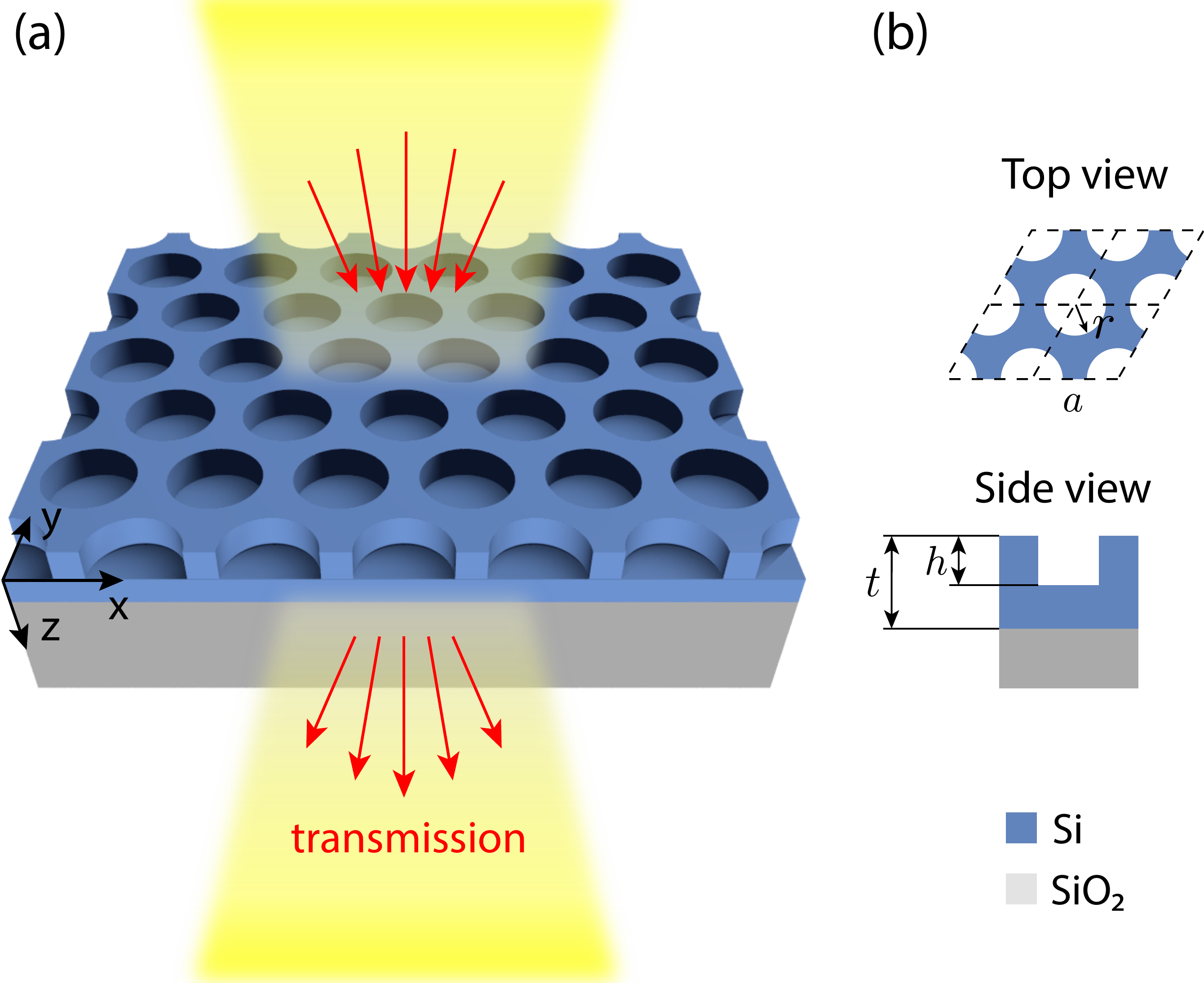}
    \caption{Schematic of the studied photonic crystal slab. A silicon layer of thickness $t=300$~nm on a $\mathrm{SiO_2}$ substrate ($n_{\mathrm{SiO_2}}=1.45$) is perforated with a hexagonal lattice of cylindrical holes (period $a=600$~nm, depth $h=235$~nm, radius $r=120$~nm). The structure is naturally split into the upper (grating) and lower (unpatterned) parts used in the resonant subspace approach.}
    \label{fig:fig4}
\end{figure}

Now that the theoretical framework is complete, we apply it to a representative photonic crystal slab and demonstrate its capabilities in practice. As a testbed we consider a silicon layer of thickness $t=300$~nm placed on a $\mathrm{SiO_2}$ substrate and perforated with a hexagonal lattice of cylindrical holes (see Fig.~\ref{fig:fig4}). We deliberately choose a strong grating with pronounced index contrast, which supports several closely spaced and strongly hybridized quasiguided modes near the $\Gamma$ point. This is a demanding rather than a favourable scenario for a resonant approximation, and it therefore allows us to probe the robustness of the developed approach under conditions where weak coupling approximations would fail.
Throughout this section all reference computations are performed with the FMM using $N=91$ Fourier harmonics, and the resonant subspace is restricted to $n_\mathrm{res}=10$ eigenstates. Although only three strongly coupled modes near the $\Gamma$ point are demonstrated here, we intentionally retain a somewhat larger resonant subspace: for a strong grating with heavily hybridized modes it is advantageous to keep a few additional states so that their influence on the modes of interest is also taken into account. In many simpler cases—in particular for weak gratings—it is sufficient to take $n_\mathrm{res}$ equal to the number of modes of interest or only slightly larger, which further increases the speedup.

\subsubsection{Eigenmode dispersion}
\begin{figure*}
    \centering
    \includegraphics[width=\linewidth]{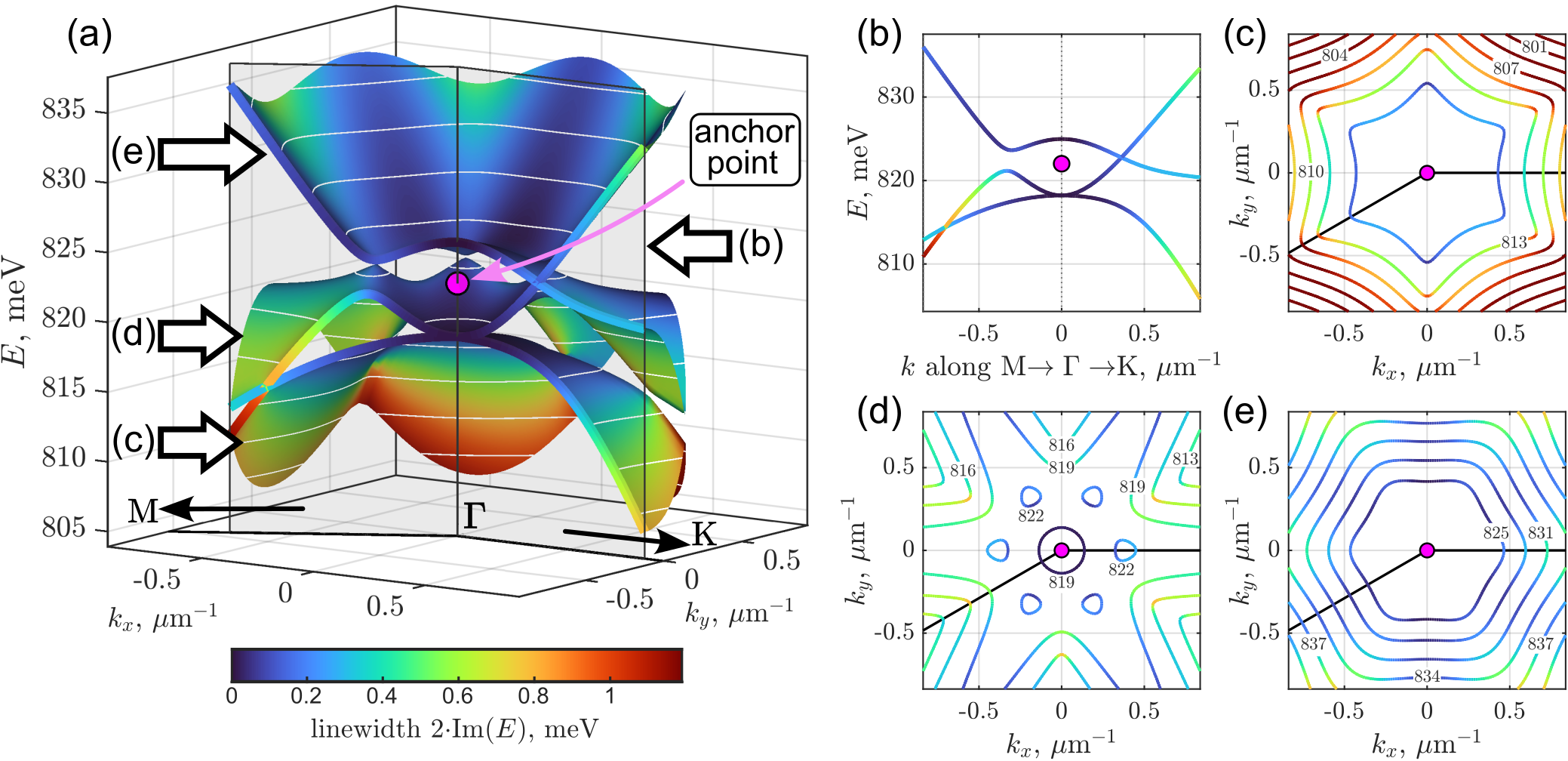}
    \caption{Eigenmode dispersion of the strong hexagonal grating obtained with the resonant subspace approach. (a) Energies of the three hybridized modes near the $\Gamma$ point as functions of the in-plane wavevector; the surface colour encodes the modal linewidth $2\,\mathrm{Im}(E)$ (inverse quality factor), and white lines are isofrequency contours. The translucent planes mark the $\Gamma\to\mathrm{M}$ and $\Gamma\to\mathrm{K}$ directions and delimit the cut-out sector. (b) Dispersion along the $\mathrm{M}\to\Gamma\to\mathrm{K}$ path, showing that inter-mode coupling varies strongly between anticrossings. (c--e) Isofrequency contours of the three modes, which reproduce the $C_6$ symmetry of the lattice and reveal hexagonal-, star-, and hyperbolic-like shapes as well as closed loops away from $\Gamma$. The whole surface is reconstructed from only four rigorous computations (anchor point plus three neighbours).}
    \label{fig:fig5}
\end{figure*}

We begin with the eigenmode dispersion, which follows directly from the effective Hamiltonian introduced in the previous section. Figure~\ref{fig:fig5}(a) shows the energies of the three hybridized modes near the $\Gamma$ point as functions of the in-plane wavevector $\mathbf{k}_\parallel=(k_x,k_y)$. We stress that the entire three-dimensional dispersion surface is reconstructed from only \emph{four} rigorous computations: one at the anchor point (located at $\Gamma$) and three in its immediate vicinity, needed to estimate the derivatives of the restricted phase matrix with respect to frequency and the two components of $\mathbf{k}_\parallel$. Performing a handful of accurate FMM computations is almost always affordable: it typically takes a fraction of a second to a few seconds, and even for very demanding structures rarely exceeds minutes. Once these anchor data are obtained, the method reduces the problem to a small number of resonances and quantities of interest, so that the subsequent evaluation at each new point amounts to the diagonalization of an $n_\mathrm{res}\times n_\mathrm{res}$ matrix and becomes essentially instantaneous. As a result, the tens and hundreds of thousands of points required to render maps such as Fig.~\ref{fig:fig5}(a) are computed on an ordinary laptop in seconds to a few minutes. In practice this means that such calculations no longer constitute a bottleneck and cease to limit the exploration or design of these structures.

The approach yields a genuinely complete description of the resonances. Since the eigenvalues of the effective Hamiltonian are complex, the imaginary part of the energy—i.e. the resonance linewidth $2\,\mathrm{Im}(E)$, or equivalently the inverse quality factor—is obtained automatically together with the dispersion and is shown as the surface colour in Fig.~\ref{fig:fig5}. From the same underlying data one can freely construct the dispersion in a multidimensional parameter space (Fig.~\ref{fig:fig5}~(a)), cuts along arbitrary directions (Fig.~\ref{fig:fig5}~(b)), or isofrequency contours (Fig.~\ref{fig:fig5}~(c--e)) without any additional rigorous computations. The isofrequency contours faithfully reproduce the $C_6$ symmetry of the underlying crystal and provide a transparent picture of the dispersion: we observe hexagonal-, star-, and hyperbolic-like contours, and even closed loops appearing away from the $\Gamma$ point. They also make it evident that the quality factor of a mode may vary strongly along a given isofrequency line, so that states of the same energy can exhibit markedly different linewidths depending on their position in reciprocal space. The dispersion cut in Fig.~\ref{fig:fig5}~(b) further illustrates the intricate coupling landscape: at some anticrossings the interaction between the modes is substantial, whereas at others it is barely noticeable, and one can directly follow how the modes hybridize, where their quality factor is enhanced through interaction, and how closed dispersion pockets emerge and disappear.

It is worth emphasizing that the computational gain originates from the strong reduction of the matrix size. Because both matrix inversion and spectral decomposition scale as the cube of the matrix dimension, and because $n_\mathrm{res}\ll N$, the acceleration is very large even in the present, deliberately unfavourable case, and it becomes substantially larger for typical problems and weak gratings. In fact, the evaluation is so fast that dedicated optimization of this stage is hardly warranted. The attainable speedup is effectively set by the number of harmonics used in the reference calculations. Finally, we note that high-$Q$ resonances pose no difficulty whatsoever for the method: they are simply modes with a small imaginary part and are treated as accurately as any others, whereas their straightforward resolution by conventional means would require sampling the spectra on prohibitively dense grids.

\subsubsection{Transmission spectra}
\begin{figure*}
    \centering
    \includegraphics[width=\linewidth]{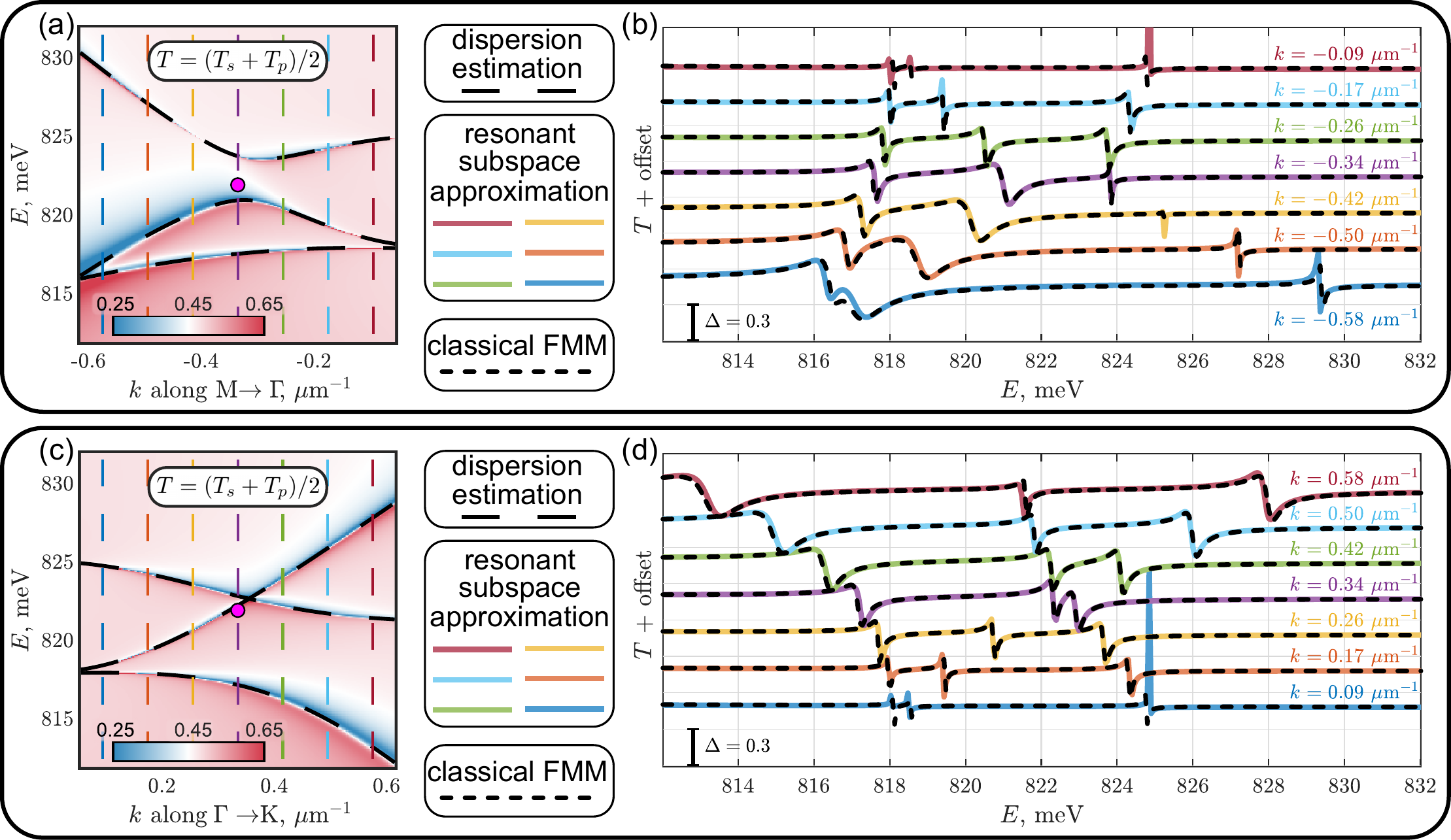}
    \caption{Transmission spectra of the hexagonal grating computed with the resonant subspace approach. (a,c) Maps of the unpolarized transmittance $T=(T_s+T_p)/2$ along the $\mathrm{M}\to\Gamma$ and $\Gamma\to\mathrm{K}$ directions, each based on its own anchor point (pink marker). Black dashed lines show the dispersion predicted by the corresponding effective Hamiltonian; it coincides with the spectral resonances since both are obtained within the same approximation. Vertical dashed lines mark the constant-$k$ cuts compared in (b,d) against direct FMM computations ($N=91$). The agreement is good along all cuts; it is nearly perfect at the wavevector of the anchor point (purple) and degrades gradually with increasing distance from it.}
    \label{fig:fig6}
\end{figure*}

The method naturally provides not only the eigenmode dispersion but also the optical spectra of interest. To this end it suffices to include the transmission coefficients of the principal diffraction channel among the quantities of interest, after which transmission spectra can be evaluated for an arbitrary polarization. As an example, Fig.~\ref{fig:fig6}~(a,c) shows maps of the unpolarized transmittance $T=(T_s+T_p)/2$ along the $\mathrm{M}\to\Gamma$ and $\Gamma\to\mathrm{K}$ directions. In contrast to the dispersion study, here we deliberately place the two anchor points away from the $\Gamma$ point, each tailored to a more detailed investigation of the corresponding cut in $\omega$--$k$ space.

On each map the black dashed lines indicate the modal dispersion predicted by the effective Hamiltonian. As expected, these lines coincide with the spectral resonances, since both the dispersion and the spectra are obtained within the same resonant approximation. To assess the reliability of the results, we compare them with rigorous FMM computations performed with $N=91$ harmonics. Such computations are far more expensive, and generating an entire spectral map in this way would be prohibitively time-consuming. We therefore select seven constant-$k$ cuts for each map and present the comparison in Fig.~\ref{fig:fig6}~(b,d), respectively. The agreement is rather good along all cuts: the resonant subspace approach reproduces not only the overall spectral background but also the resonant line shapes and the full pattern of mode hybridization. As anticipated for an extrapolation from a single anchor point, the correspondence is almost perfect at the wavevector matching the anchor point (purple curves) and deteriorates only gradually as one moves away from it. Even so, within the considered range the approximation remains not merely qualitative but quantitatively accurate, while being obtained incomparably faster. If a particular region requires higher fidelity, it can be studied either by placing an additional anchor point at its centre or by resorting to conventional methods within an already narrow window.

\subsubsection{Dependence on geometrical parameters}
\begin{figure*}
    \centering
    \includegraphics[width=\linewidth]{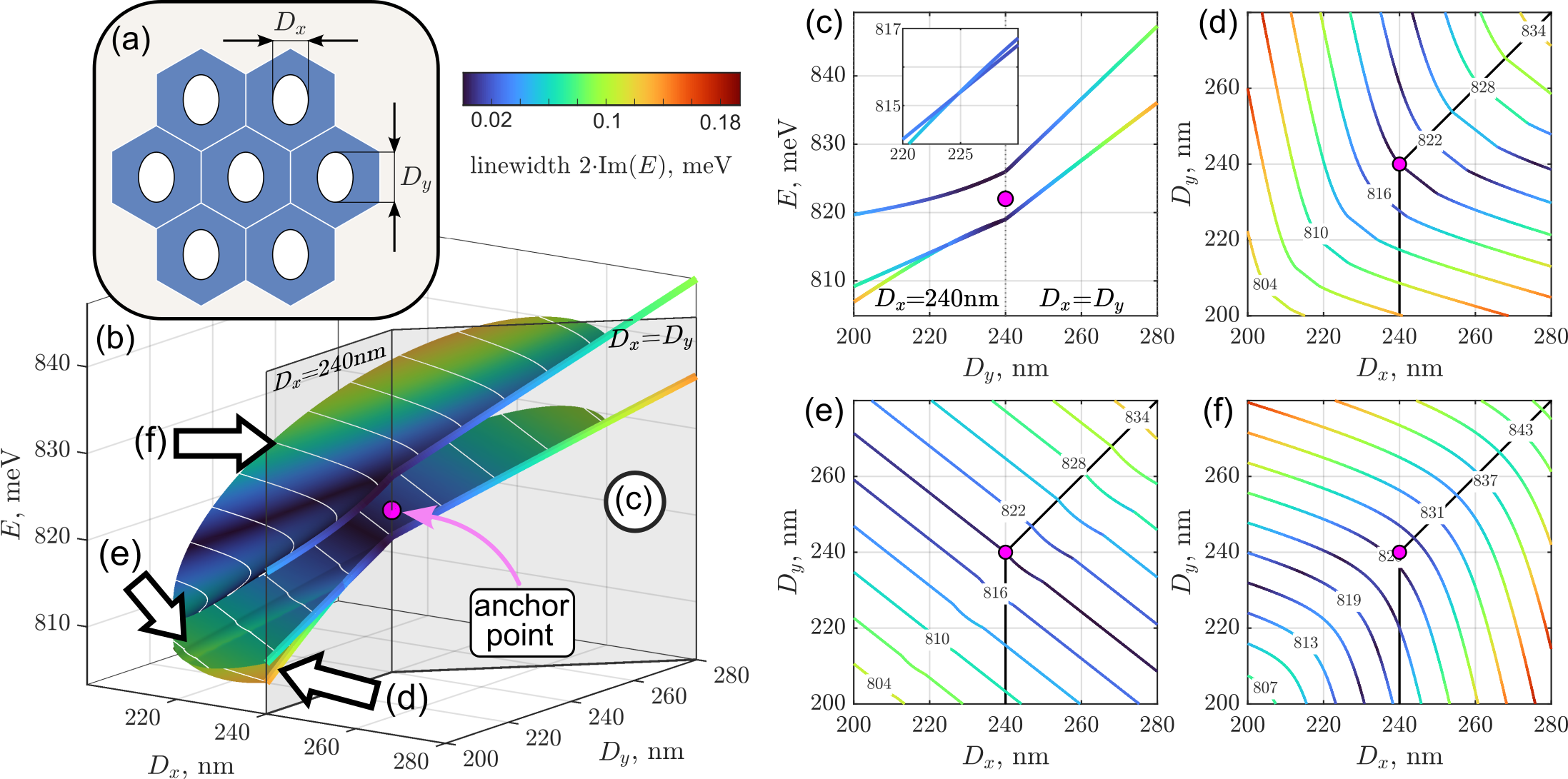}
    \caption{Eigenmodes as functions of the elliptic-hole diameters $D_x$ and $D_y$. (a) Unit cell with the elliptic hole. (b) Energies of the three modes versus $D_x$ and $D_y$; the surface colour encodes the linewidth $2\,\mathrm{Im}(E)$, white lines are isofrequency contours, and the translucent planes mark the $D_x=D_y$ and $D_x=240$~nm directions. (c) Dispersion along these directions, showing the lifting of the degeneracy of the two lower modes for $D_x\neq D_y$ and a fine anticrossing near $D_y\approx225$~nm (inset). (d--f) Isofrequency contours of the three modes. The mode energies depend primarily on the sum $D_x+D_y$ and only weakly on the difference $D_x-D_y$; the contours demonstrate a number of fine features: opposite bending of modes~1 and~3, the nearly flat contours of mode~2 with a faint asymmetric ``hat'' for $D_x\gtrsim D_y$, and the linewidth change upon symmetry breaking.}
    \label{fig:fig7}
\end{figure*}

Finally, we emphasize that the parameters $\mathbf{p}$ are not restricted to the frequency and wavevector: the very same formalism describes the dependence of the resonances on any structural parameter on an equal footing. To illustrate this, we deform the circular holes into ellipses with in-plane diameters $D_x$ and $D_y$ (see Fig.~\ref{fig:fig7}~(a)) and study how the symmetry breaking of the meta-atom affects the modes.

The reconstructed energies of the three modes as functions of $D_x$ and $D_y$ are shown in Fig.~\ref{fig:fig7}~(b), with the corresponding cuts and isofrequency contours in panels~(c) and~(d--f). To leading order the mode energies are governed by the total hole size, i.e. by the sum $D_x+D_y$, and depend only weakly on the difference $D_x-D_y$. Breaking the symmetry of the hole ($D_x\neq D_y$) lifts the degeneracy of the two lower modes, as clearly seen in the dispersion cut of Fig.~\ref{fig:fig7}~(c). On top of this overall behaviour, the map reveals remarkably fine features. In particular, the isofrequency contours of the first and third modes bend in opposite directions, while those of the second mode are almost flat except for a very faint, asymmetrically located ``hat'' for $D_x\gtrsim D_y$. This hat is precisely ``cut out'' from the parabolic-like contours of the first mode and provides evidence of an anticrossing between two extremely close modes, visible as the tiny gap near $D_y\approx225$~nm in the inset of Fig.~\ref{fig:fig7}~(c). Although the practical role of such a subtle effect remains to be understood, its very detection highlights the power of the resonant subspace approach: features this delicate would be exceedingly hard to resolve, let alone reliably characterize, with a straightforward point-by-point computation.

Equally instructive is the behaviour of the linewidth. As Fig.~\ref{fig:fig7}~(f) shows, introducing an asymmetry $D_x\neq D_y$ can substantially raise or lower the quality factor of a mode at essentially the same energy. This is a clear example of how effortless access to the dependence of the modes on geometrical parameters can be exploited to engineer desirable optical properties, and it illustrates the potential of the method as a practical tool for the design and optimization of resonant photonic crystal slabs.

\section{Discussion and conclusion}
In this work we have presented an efficient computational approach for the simulation of resonant photonic crystal slabs. Its central idea is a generalization of the familiar scalar round-trip phase, acquired by light travelling up and down inside a slab waveguide, to periodic structures, in which a whole bunch of coupled Fourier harmonics propagates and the phase accordingly becomes a matrix quantity. Building on this picture, we have introduced the round-trip and phase matrices, whose eigenvalues approaching unity signal the formation of quasiguided modes, and shown how the resonant response of the structure can be reconstructed from them. The approach rests on two practical observations: (i) the eigenvectors of the resonant modes span a subspace that varies only weakly with the structural and illumination parameters, so that the corresponding projector can be fixed at a single anchor point; and (ii) the scattering matrices of the non-resonant constituents, and hence the restricted round-trip and phase matrices, are smooth functions of all parameters. Together these properties enable a simple yet accurate ``smart'' extrapolation from an anchor point that faithfully captures not only the smooth background of the spectra but also the resonant lineshapes and, most importantly, the hybridization of the interacting modes.

What primarily distinguishes our method from most existing resonant approximations is that it treats the photon energy and any other parameter—wavevector, geometric dimensions, or even the permittivity of the constituent materials—on an equal footing within a single, universal framework. Conventional schemes typically describe an eigenmode as a pole in the complex energy plane for a fixed set of the remaining parameters, which forces one to perform a separate resonant expansion at every point of the parametric space and then to stitch these expansions together while manually tracking the modes through their avoided crossings. Our formulation removes this bottleneck entirely: the fixed anchor-point basis automatically ensures a consistent labelling of the modes across the whole local region and naturally accommodates their hybridization, so that neither manual matching nor fragile automated sorting is required.

The practical consequences of this reformulation are substantial. As we have demonstrated, a careful expansion around a single anchor point requires accurate, and therefore relatively expensive, rigorous computations at only a few neighbouring points— at least one per parameter of interest, in addition to the anchor itself. Even when a large number of Fourier harmonics is needed for accuracy, these few reference calculations are almost always affordable. Afterwards, the problem is reduced to a small number of resonances and quantities of interest, and every subsequent evaluation amounts to the diagonalization of a tiny $n_\mathrm{res}\times n_\mathrm{res}$ matrix. Because both matrix inversion and spectral decomposition scale as the cube of the matrix size and $n_\mathrm{res}\ll N$, the resulting acceleration is extremely large for typical problems. In effect, reconstruction of high-resolution maps comprising typically tens of thousands of points takes mere seconds to a few minutes on an ordinary laptop. From the very same data one obtains, without any additional rigorous computations, the full dispersion in a multidimensional parameter space, cuts along arbitrary directions, isofrequency contours, and—since the eigenvalues of the effective Hamiltonian are complex—the modal linewidths, i.e. the inverse quality factors, automatically together with the mode energies. We stress that high-$Q$ resonances present no difficulty whatsoever for the method: they are simply modes with a small imaginary part and are treated as accurately as any others, whereas their straightforward resolution by conventional means would demand prohibitively dense sampling grids.

We have illustrated these capabilities on a strong, deliberately unfavourable silicon grating with several heavily hybridized modes. The reconstructed band structure preserves the $C_6$ symmetry of the underlying lattice, reveals hexagonal-, star-, and hyperbolic-like isofrequency contours and closed loops away from the $\Gamma$ point, and reproduces the transmission spectra in excellent agreement with direct FMM computations. Moreover, by treating the elliptic-hole diameters as parameters, we have effortlessly uncovered extremely fine features—a subtle anticrossing between two nearly degenerate modes and a pronounced, geometry-controlled variation of the quality factor at essentially fixed energy—that would be exceedingly hard to detect, let alone reliably characterize, with a straightforward point-by-point approach. These examples already indicate how easy access to the parametric dependence of both dispersion and linewidth can be turned into a practical tool for the design and optimization of resonant structures, where any reflection, transmission, diffraction, or absorption coefficient may be studied as a function of one or several parameters simultaneously.

At the same time, several challenges remain and may serve as cornerstones for the further development of the approach towards a universal computational package. First, extrapolation is intrinsically local, so covering a wide range of parameters requires several anchor points, each responsible for its own region; an automated placement and matching of such anchors is a natural next step. Second, Rayleigh anomalies, at which the derivatives with respect to energy and wavevector become discontinuous, break the smoothness assumption and thus call for a dedicated treatment---most probably along the lines of the resonant approximations formulated in terms of the out-of-plane wavevector $k_z$ instead of the energy~\cite{akimov2011optical,gromyko2022resonant}, which removes the corresponding branch point. Third, although linear extrapolation is the simplest choice, higher-order Taylor expansions or other suitably smooth fitting strategies may improve accuracy or extend the region of validity. Fourth, quantities other than far-field spectra—in particular local fields generated by an external source, or the emission of a local source embedded in the structure—can be described within the same approximation. Fifth, the scheme can be generalized to the hybridization of resonances originating from different layers of a multilayer stack, where it could be combined with the existing resonant mode coupling approximation~\cite{gippius2010resonant,gromyko2023resonant1,gromyko2023resonant2}, and, more broadly, beyond the scattering-matrix description of periodic slabs to other geometries such as multilayer spheres. Finally, the present derivations open the way to semi-analytical evaluation of integral characteristics such as the Purcell factor, Casimir forces, and near-field energy transfer: both the smooth background and the Lorentzian poles can be integrated analytically, obviating the need to sum over extremely dense grids that resolve narrow modes.

Altogether, the demonstrated approach is already fast, accurate, and convenient for practical calculations, and it carries high potential for further development and for a broad range of applications in the modelling, design, and optimization of resonant photonic crystal slabs.

\section{Acknowledgements}
The work was supported by the Russian Science Foundation (Grant no. 22-12-00351-$\Pi$).
\appendix
\section{Phase matrix as a logarithm of the round-trip matrix}
\label{sec:app_phase}
In the main text we have defined the phase matrix $\hat{\Phi}^\mathbf{p}_\mathrm{res}$ through the logarithms of the round-trip eigenvalues, and then claimed that the exponential relation $\mathbbm{g}^{\mathbf{p},\mathbf{p}_0}=\exp(i\hat{\phi}^{\mathbf{p},\mathbf{p}_0})$ survives both the restriction to the resonant subspace and the transition to the fixed basis of the anchor point. Here we derive this relation starting directly from the definitions and specify the conditions under which it holds.

It is convenient to collect the resonant right and left eigenvectors of the round-trip matrix at the point $\mathbf{p}$ into the rectangular $2N\times n_\mathrm{res}$ matrices $\mathbb{V}^{\mathbf{p}}_\mathrm{res}=(\ket{\mathbf{V}^\mathbf{p}_i})_{i\in\mathcal{I}}$ and $\mathbb{W}^{\mathbf{p}}_\mathrm{res}=(\ket{\mathbf{W}^\mathbf{p}_i})_{i\in\mathcal{I}}$. The biorthonormality relation $\braket{\mathbf{W}^\mathbf{p}_i|\mathbf{V}^\mathbf{p}_j}=\delta_{ij}$ then splits into two statements, which should not be confused with each other:
\begin{equation}
    \mathbb{W}^{\dagger,\mathbf{p}}_\mathrm{res}\mathbb{V}^{\mathbf{p}}_\mathrm{res} = \hat{I}_{n_\mathrm{res}},
    \qquad
    \mathbb{V}^{\mathbf{p}}_\mathrm{res}\mathbb{W}^{\dagger,\mathbf{p}}_\mathrm{res} = \mathbb{P}^{\mathbf{p}}_\mathrm{res}.
    \label{eq:app_biorth}
\end{equation}
The first product is the small $n_\mathrm{res}\times n_\mathrm{res}$ identity matrix, whereas the second one is the large $2N\times2N$ projector onto the resonant subspace, which is idempotent, $(\mathbb{P}^{\mathbf{p}}_\mathrm{res})^2=\mathbb{P}^{\mathbf{p}}_\mathrm{res}$, but is by no means the identity. These two relations are essentially the only ingredients of the derivation below. In the same notation, the resonant parts of the round-trip and phase matrices read
\begin{align}
    \mathbb{G}^{\mathbf{p}}_\mathrm{res} &= \mathbb{V}^{\mathbf{p}}_\mathrm{res}\left[\rho^\mathbf{p}\right]\mathbb{W}^{\dagger,\mathbf{p}}_\mathrm{res},\\
    \hat{\Phi}^{\mathbf{p}}_\mathrm{res} &= \mathbb{V}^{\mathbf{p}}_\mathrm{res}\left[-i\ln\rho^\mathbf{p}\right]\mathbb{W}^{\dagger,\mathbf{p}}_\mathrm{res},
    \label{eq:app_G_Phi}
\end{align}
where $[\rho^\mathbf{p}]=\operatorname{diag}(\rho^\mathbf{p}_i)\big|_{i\in\mathcal{I}}$ and $[\ln\rho^\mathbf{p}]=\operatorname{diag}(\ln\rho^\mathbf{p}_i)\big|_{i\in\mathcal{I}}$ are small diagonal matrices. The logarithm is well defined here, since the resonant eigenvalues are close to unity and, in particular, non-zero.

Let us first apply the exponential to the phase operator as it stands, in the full $2N$-dimensional space. Owing to the first of the relations~\eqref{eq:app_biorth}, each power of the phase operator retains the same outer structure,
\begin{equation}
    \left(i\hat{\Phi}^{\mathbf{p}}_\mathrm{res}\right)^n = \mathbb{V}^{\mathbf{p}}_\mathrm{res}\left[\ln\rho^\mathbf{p}\right]^n\mathbb{W}^{\dagger,\mathbf{p}}_\mathrm{res},\qquad n\geq1,
\end{equation}
so that the series might be summed term by term:
\begin{multline}
    \exp\left(i\hat{\Phi}^{\mathbf{p}}_\mathrm{res}\right)
    = \hat{I}_{2N}+\sum_{n=1}^{\infty}\frac{1}{n!}\mathbb{V}^{\mathbf{p}}_\mathrm{res}\left[\ln\rho^\mathbf{p}\right]^n\mathbb{W}^{\dagger,\mathbf{p}}_\mathrm{res}\\
    = \hat{I}_{2N}+\mathbb{V}^{\mathbf{p}}_\mathrm{res}\left[\rho^\mathbf{p}-1\right]\mathbb{W}^{\dagger,\mathbf{p}}_\mathrm{res}
    = \hat{I}_{2N}-\mathbb{P}^{\mathbf{p}}_\mathrm{res}+\mathbb{G}^{\mathbf{p}}_\mathrm{res}.
    \label{eq:app_exp_full}
\end{multline}
This result is instructive: the exponential of the phase operator does not reproduce the round-trip operator, because the zeroth term of the series lives in the whole space and returns the identity on the non-resonant complement, where $\hat{\Phi}_\mathrm{res}$ itself vanishes. The two operators do coincide on the resonant subspace though,
\begin{equation}
    \mathbb{G}^{\mathbf{p}}_\mathrm{res} = \mathbb{P}^{\mathbf{p}}_\mathrm{res}\exp\left(i\hat{\Phi}^{\mathbf{p}}_\mathrm{res}\right)\mathbb{P}^{\mathbf{p}}_\mathrm{res},
\end{equation}
and this is exactly the reason why the relation of our interest should be formulated for the small restricted matrices rather than for the large operators.

We now pass to the fixed basis of the anchor point. The restriction of an operator to the resonant subspace expressed in this basis amounts to sandwiching it between $\mathbb{W}^{\dagger,\mathbf{p}_0}_\mathrm{res}$ and $\mathbb{V}^{\mathbf{p}_0}_\mathrm{res}$, which for the two matrices of interest gives
\begin{align}
    \hat{\phi}^{\mathbf{p},\mathbf{p}_0} &= \mathbb{W}^{\dagger,\mathbf{p}_0}_\mathrm{res}\hat{\Phi}^{\mathbf{p}}_\mathrm{res}\mathbb{V}^{\mathbf{p}_0}_\mathrm{res} = \hat{t}\left[-i\ln\rho^\mathbf{p}\right]\hat{s},
    \label{eq:app_phi_ts}\\
    \mathbbm{g}^{\mathbf{p},\mathbf{p}_0} &= \mathbb{W}^{\dagger,\mathbf{p}_0}_\mathrm{res}\mathbb{G}^{\mathbf{p}}\mathbb{V}^{\mathbf{p}_0}_\mathrm{res} \approx \hat{t}\left[\rho^\mathbf{p}\right]\hat{s},
    \label{eq:app_g_ts}
\end{align}
where we have introduced the small overlap matrices between the current and the anchor bases,
\begin{equation}
    \hat{t} = \mathbb{W}^{\dagger,\mathbf{p}_0}_\mathrm{res}\mathbb{V}^{\mathbf{p}}_\mathrm{res},
    \qquad
    \hat{s} = \mathbb{W}^{\dagger,\mathbf{p}}_\mathrm{res}\mathbb{V}^{\mathbf{p}_0}_\mathrm{res},
\end{equation}
with the elements $t_{ij}=\braket{\mathbf{W}^{\mathbf{p}_0}_i|\mathbf{V}^{\mathbf{p}}_j}$ and $s_{ij}=\braket{\mathbf{W}^{\mathbf{p}}_i|\mathbf{V}^{\mathbf{p}_0}_j}$. The approximate sign in Eq.~\eqref{eq:app_g_ts} is a matter of bookkeeping rather than of principle. The phase matrix is defined through $\hat{\Phi}^{\mathbf{p}}_\mathrm{res}$, which is confined to the resonant subspace by construction, so that substituting its spectral form \eqref{eq:app_G_Phi} makes Eq.~\eqref{eq:app_phi_ts} an identity, whereas $\mathbbm{g}^{\mathbf{p},\mathbf{p}_0}$ is defined through the \textit{full} matrix $\mathbb{G}^{\mathbf{p}}$, so that discarding its non-resonant part amounts to $\mathbb{P}^{\mathbf{p}_0}_\mathrm{res}\mathbb{G}^{\mathbf{p}}\mathbb{P}^{\mathbf{p}_0}_\mathrm{res}\approx\mathbb{P}^{\mathbf{p}}_\mathrm{res}\mathbb{G}^{\mathbf{p}}\mathbb{P}^{\mathbf{p}}_\mathrm{res}$. Since this is the nearly constant projector approximation already adopted in the main text, the two definitions are interchangeable within the accuracy of the approach, and neither of them is preferred.

The very same approximation provides the key property of the overlap matrices. Indeed,
\begin{equation}
    \hat{s}\hat{t}
    = \mathbb{W}^{\dagger,\mathbf{p}}_\mathrm{res}\mathbb{P}^{\mathbf{p}_0}_\mathrm{res}\mathbb{V}^{\mathbf{p}}_\mathrm{res}
    \approx \mathbb{W}^{\dagger,\mathbf{p}}_\mathrm{res}\mathbb{P}^{\mathbf{p}}_\mathrm{res}\mathbb{V}^{\mathbf{p}}_\mathrm{res}
    = \hat{I}_{n_\mathrm{res}},
    \label{eq:app_st}
\end{equation}
and, since $\hat{s}$ and $\hat{t}$ are square, this immediately implies $\hat{t}\hat{s}=\hat{I}_{n_\mathrm{res}}$ as well, so that $\hat{s}=\hat{t}^{-1}$. Physically, Eq.~\eqref{eq:app_st} states that the resonant eigenvectors at the point $\mathbf{p}$ are exhaustively expandable over the anchor-point ones, $\ket{\mathbf{V}^{\mathbf{p}}_j}=\sum_{i\in\mathcal{I}}\ket{\mathbf{V}^{\mathbf{p}_0}_i}t_{ij}$, and that this expansion is invertible --- which is precisely what a nearly constant resonant subspace means.

Therefore, the restricted phase and round-trip matrices are brought to their diagonal forms by one and the same similarity transformation,
\begin{equation}
    \hat{\phi}^{\mathbf{p},\mathbf{p}_0} = \hat{t}\left[-i\ln\rho^\mathbf{p}\right]\hat{t}^{-1},
    \qquad
    \mathbbm{g}^{\mathbf{p},\mathbf{p}_0} = \hat{t}\left[\rho^\mathbf{p}\right]\hat{t}^{-1},
    \label{eq:app_similarity}
\end{equation}
and the desired relation follows in a couple of lines:
\begin{multline}
    \exp\left(i\hat{\phi}^{\mathbf{p},\mathbf{p}_0}\right)
    = \sum_{n=0}^{\infty}\frac{1}{n!}\left(\hat{t}\left[\ln\rho^\mathbf{p}\right]\hat{t}^{-1}\right)^n\\
    = \hat{t}\left(\sum_{n=0}^{\infty}\frac{\left[\ln\rho^\mathbf{p}\right]^n}{n!}\right)\hat{t}^{-1}
    = \hat{t}\left[\rho^\mathbf{p}\right]\hat{t}^{-1}
    = \mathbbm{g}^{\mathbf{p},\mathbf{p}_0},
\end{multline}
where the intermediate factors $\hat{t}^{-1}\hat{t}$ cancel in every term of the series and the exponential of a diagonal matrix is taken elementwise, $\exp[\ln\rho^\mathbf{p}]=[\rho^\mathbf{p}]$. In contrast to Eq.~\eqref{eq:app_exp_full}, the $n=0$ term now correctly reproduces the identity, because the restriction of the projector to the resonant subspace is the small identity matrix $\hat{I}_{n_\mathrm{res}}$ and not $\hat{I}_{2N}$ [see the first of the relations~\eqref{eq:app_biorth}].

Several consequences are worth mentioning. First, Eq.~\eqref{eq:app_similarity} shows that $\hat{\phi}^{\mathbf{p},\mathbf{p}_0}$ and $\mathbbm{g}^{\mathbf{p},\mathbf{p}_0}$ share their eigenvectors, which are the columns of $\hat{t}$, while their eigenvalues are related as $\tilde{\rho}^\mathbf{p}_m=e^{i\phi^\mathbf{p}_m}$, as stated in the main text. This also endows the eigenvector matrix $\hat{v}^\mathbf{p}$ with a transparent meaning: up to the normalization of its columns, it is an estimate of $\hat{t}$, that is, of the decomposition of the true resonant eigenvectors at the point $\mathbf{p}$ over the anchor-point basis.

Second, at the anchor point itself the derivation is exact: there $\hat{t}=\hat{s}=\hat{I}_{n_\mathrm{res}}$, so that $\hat{\phi}^{\mathbf{p}_0,\mathbf{p}_0}=[-i\ln\rho^{\mathbf{p}_0}]$ and $\mathbbm{g}^{\mathbf{p}_0,\mathbf{p}_0}=[\rho^{\mathbf{p}_0}]$ are both diagonal and trivially related by the exponential. Away from the anchor point, the accuracy of the relation is controlled by the deviation of the product $\hat{s}\hat{t}=\hat{I}_{n_\mathrm{res}}-\mathbb{W}^{\dagger,\mathbf{p}}_\mathrm{res}\left(\mathbb{P}^{\mathbf{p}}_\mathrm{res}-\mathbb{P}^{\mathbf{p}_0}_\mathrm{res}\right)\mathbb{V}^{\mathbf{p}}_\mathrm{res}$ from the identity, and thus vanishes together with the variation of the resonant projector; no assumption beyond the one already made in the main text is introduced here.

Third, since the resonant eigenvalues are non-zero, the matrix $\mathbbm{g}^{\mathbf{p},\mathbf{p}_0}$ is non-singular and the proven relation might be inverted,
\begin{equation}
    \hat{\phi}^{\mathbf{p},\mathbf{p}_0} = -i\ln\mathbbm{g}^{\mathbf{p},\mathbf{p}_0},
\end{equation}
which is namely the form used in practice (see Fig.~\ref{fig:fig2}~(d,f)): the restricted round-trip matrix is obtained by a straightforward projection of $\mathbb{G}^{\mathbf{p}}$ onto the anchor basis, and the phase matrix follows from its matrix logarithm, without any need to diagonalize the large operators at the point $\mathbf{p}$. Read as a definition of $\hat{\phi}^{\mathbf{p},\mathbf{p}_0}$, this relation is exact by construction, which is one more reason why the choice between the two routes is immaterial.

Finally, the logarithm is a multivalued function, which makes the phase matrix non-uniquely defined as well: the substitution $\ln\rho^\mathbf{p}_m\to\ln\rho^\mathbf{p}_m+2\pi i n_m$ with arbitrary integers $n_m$ shifts it by $2\pi\sum_m n_m\hat{t}\hat{e}_{mm}\hat{t}^{-1}$, where $\hat{e}_{mm}$ is the corresponding diagonal unit matrix. The relation proven above holds for any such choice, since all the branches become indistinguishable after the exponentiation, so that the round-trip eigenvalues and all the physical results are insensitive to it. The branch does matter for the extrapolation, though: a shifted phase matrix is equally valid but is no longer a smooth continuation of the one defined at the anchor point. We therefore apply the standard natural logarithm definition at the anchor point and further expand it continuously in parametric space, which is exactly what makes the phase matrix a slowly varying, nearly linear function of the parameters.
\section{Equivalent formulation through the downward round-trip matrix}
\label{sec:app_dd}
The splitting of the structure into the upper and the lower parts leaves one more freedom: the multiple-reflection series in the section between the two parts might be summed either for the up-going or for the down-going wave. Denoting the corresponding amplitudes in this section by $\mathbf{u}$ and $\mathbf{d}$, and the external incoming ones by $\mathbf{a}^\downarrow$ and $\mathbf{a}^\uparrow$, we have the self-consistency relations
\begin{equation}
    \mathbf{d} = \mathbb{S}_1^{\downarrow\downarrow}\mathbf{a}^\downarrow+\mathbb{S}_1^{\downarrow\uparrow}\mathbf{u},
    \qquad
    \mathbf{u} = \mathbb{S}_2^{\uparrow\downarrow}\mathbf{d}+\mathbb{S}_2^{\uparrow\uparrow}\mathbf{a}^\uparrow,
    \label{eq:app_ud}
\end{equation}
which, resolved with respect to $\mathbf{u}$, give Eqs.~\eqref{Eq:Ss_DAA}--\eqref{eq:DAA} of the main text. Resolving them with respect to $\mathbf{d}$ instead, one arrives at the equivalent formulation
\begin{multline}
    \mathbb{S}=
    \begin{pmatrix}
        \hat{0} & \mathbb{S}_2^{\downarrow\uparrow}\\
        \mathbb{S}_1^{\uparrow\downarrow} & \mathbb{S}_1^{\uparrow\uparrow}\mathbb{S}_2^{\uparrow\uparrow}
    \end{pmatrix}
    +\\
    \begin{pmatrix}
        \mathbb{S}_2^{\downarrow\downarrow}\\
        \mathbb{S}_1^{\uparrow\uparrow}\mathbb{S}_2^{\uparrow\downarrow}
    \end{pmatrix}
    \mathbb{D}^{\downarrow\downarrow}
    \begin{pmatrix}
        \mathbb{S}_1^{\downarrow\downarrow} & \mathbb{S}_1^{\downarrow\uparrow}\mathbb{S}_2^{\uparrow\uparrow}
    \end{pmatrix},
    \label{eq:app_Ss_DDD}
\end{multline}
where the down-going round-trip and denominator matrices differ from their up-going counterparts only by the order of the two reflection matrices,
\begin{equation}
    \mathbb{G}^{\downarrow\downarrow} = \mathbb{S}_1^{\downarrow\uparrow}\mathbb{S}_2^{\uparrow\downarrow},
    \qquad
    \mathbb{D}^{\downarrow\downarrow} = \left(\hat{I}-\mathbb{G}^{\downarrow\downarrow}\right)^{-1},
\end{equation}
since the round trip is now started from the down-going wave.

The equivalence of the two formulations is provided by the push-through identity
\begin{equation}
    \mathbb{D}^{\downarrow\downarrow}\mathbb{S}_1^{\downarrow\uparrow} = \mathbb{S}_1^{\downarrow\uparrow}\mathbb{D}^{\uparrow\uparrow},
    \qquad
    \mathbb{D}^{\downarrow\downarrow} = \hat{I}+\mathbb{S}_1^{\downarrow\uparrow}\mathbb{D}^{\uparrow\uparrow}\mathbb{S}_2^{\uparrow\downarrow},
    \label{eq:app_pushthrough}
\end{equation}
the first of which is verified by multiplying it by $(\hat{I}-\mathbb{G}^{\downarrow\downarrow})$ from the left and by $(\hat{I}-\mathbb{G}^{\uparrow\uparrow})$ from the right, which turns it into the trivial equality $\mathbb{S}_1^{\downarrow\uparrow}-\mathbb{S}_1^{\downarrow\uparrow}\mathbb{S}_2^{\uparrow\downarrow}\mathbb{S}_1^{\downarrow\uparrow}=\mathbb{S}_1^{\downarrow\uparrow}-\mathbb{S}_1^{\downarrow\uparrow}\mathbb{S}_2^{\uparrow\downarrow}\mathbb{S}_1^{\downarrow\uparrow}$. Both Eq.~\eqref{Eq:Ss_DAA} and Eq.~\eqref{eq:app_Ss_DDD} are thus particular ways of writing the same Redheffer star product $\mathbb{S}=\mathbb{S}_1\otimes\mathbb{S}_2$. The same pair of Fabry--P\'{e}rot operators, related by the very same identity, appears in the resonant mode coupling approximation of stacked structures~\cite{gromyko2023resonant1}.

Finally, $\mathbb{G}^{\uparrow\uparrow}$ and $\mathbb{G}^{\downarrow\downarrow}$ are the two products of the same pair of matrices taken in the opposite order and therefore share the characteristic polynomial, so that the resonance condition $\rho_i\to1$ is the same for both propagation directions, as it should be.

\bibliography{sample}

\end{document}